\documentclass[twocolumn, usenatbib,useAMS,fleqn,usenames,dvipsnames]{aastex63}

\usepackage{graphicx} 
\graphicspath{{./figures/}} 

\usepackage{amsmath,amsfonts,amsthm,bm}
\usepackage{enumitem}
\usepackage{float}

\begin{document}

\title{Regulation of Star Formation by a Hot Circumgalactic Medium}

\author[0000-0002-5840-0424]{Christopher Carr}
\affiliation{Department of Astronomy, Columbia University, 550 West 120th Street, New York, NY, 10027, USA}

\author[0000-0003-2630-9228]{Greg L. Bryan}
\affiliation{Department of Astronomy, Columbia University, 550 West 120th Street, New York, NY, 10027, USA}
\affiliation{Center for Computational Astrophysics, Flatiron Institute, New York, NY 10010, USA}

\author[0000-0003-3806-8548]{Drummond B. Fielding}
\affiliation{Center for Computational Astrophysics, Flatiron Institute, New York, NY 10010, USA}

\author[0000-0002-2499-9205]{Viraj Pandya}
\altaffiliation{Hubble Fellow}
\affiliation{Columbia Astrophysics Laboratory, Columbia University, 550 West 120th Street, New York, NY 10027, USA}
\affiliation{Center for Computational Astrophysics, Flatiron Institute, New York, NY 10010, USA}

\author[0000-0002-6748-6821]{Rachel S. Somerville}
\affiliation{Center for Computational Astrophysics, Flatiron Institute, New York, NY 10010, USA}

\begin{abstract}
Galactic outflows driven by supernovae (SNe) are thought to be a powerful regulator of a galaxy's star-forming efficiency. Mass, energy, and metal outflows ($\eta_M$, $\eta_E$, and $\eta_Z$, here normalized by the star formation rate, the SNe energy and metal production rates, respectively) shape galaxy properties by both ejecting gas and metals out of the galaxy and by heating the circumgalactic medium (CGM), preventing future accretion. Traditionally, models have assumed that galaxies self-regulate by ejecting a large fraction of the gas which enters the interstellar medium (ISM), even though such high mass-loadings are in growing tension with observations. To better understand how the relative importance of ejective (i.e. high mass-loading) vs preventative (i.e. high energy-loading) feedback affects the present-day properties of galaxies, we develop a simple gas-regulator model of galaxy evolution, where the stellar mass, ISM, and CGM are modeled as distinct reservoirs which exchange mass, metals, and energy at different rates within a growing halo. Focusing on the halo mass range from $10^{10}$ to $10^{12} M_{\odot}$, we demonstrate that, with reasonable parameter choices, we can reproduce the stellar-to-halo mass relation and the ISM-to-stellar mass relation with low mass-loaded ($\eta_M \sim 0.1-10$) but high energy-loaded ($\eta_E \sim 0.1-1$) winds, with self-regulation occurring primarily through heating and cooling of the CGM. We show that the model predictions are robust against changes to the mass-loading of outflows but are quite sensitive to our choice of the energy-loading, preferring $\eta_E \sim 1$ for the lowest mass halos and $\sim 0.1$ for Milky Way-like halos.  
\end{abstract}

\keywords{Circumgalactic medium (1879), Galactic winds (572), Galaxies (573), Galaxy evolution (594), Galaxy physics (612), Galactic and Extragalactic astronomy (563)}

\section{Introduction} \label{sec:intro}
Early theories of galaxy formation proposed that galaxies form from the cooling gas that sinks to the centers of virialized dark matter halos and self-gravitates to form stars \citep{white_core_1978}. However, early models ran into the longstanding problem of ``overcooling", where the accretion of gas in halos form galaxies that are too efficient at producing stars, particularly in low-mass and high-mass halos \citep{dekel_origin_1986,white_galaxy_1991}. This manifests as a mismatch in the stellar-to-halo mass relation, which compares the typical stellar masses in galaxies in the real universe to their respective dark matter halos via abundance matching techniques \citep{Moster2010, 2019_behroozi}.

Later generations of galaxy formation modeling have attempted to address this problem of overcooling by introducing feedback from supernova (SNe) explosions to make star formation less efficient in Milky Way and lower-mass halos \citep{somerville_physical_2015}. Massive stars, upon their death, deposit enormous quantities of energy, enriched gas, and momentum into the surrounding interstellar medium (ISM), regulating star formation by providing a source of effective pressure that counterbalances the local disk weight, supporting the disk against gravitational collapse \citep[e.g.,][]{Thompson2005,ostriker2022}. However, feedback not only regulates star formation within the ISM, by introducing a source of heating and turbulence \citep{krumholz_turbulence}, but also through its ability to drive large-scale multi-phase outflows with mass loss rates comparable to the galaxy's star formation rate, quantified by the mass-loading factor $\eta_M$ \citep{heckman_nature_1990,strickland_starburst-driven_2000, strickland_supernova_2009}. The presence of these galactic winds has been confirmed by observations of outflows in star-forming galaxies at high and low redshift \citep[for reviews see][]{veilleux_galactic_2005, rupke_review_2018}. Once these outflows leave the ISM, their mass, energy, and metals are either deposited into the circumgalactic medium (CGM) \citep{Heitsch2009}, recycled and accreted back onto the disk in a 'galactic fountain' \citep{Shapiro1976, Bregman1980, Marasco2022a}, or even driven out of the halo entirely \citep{oppenheimer_mass_2008, keres_galaxies_2009, 2010MNRAS.406.2325O}. SNe feedback, which is the focus of this work, is important but insufficient to explain the observed star formation inefficiencies and outflows in halos with masses $\gtrsim 10^{12} M_{\odot}$, and thus requires additional feedback from active galactic nuclei (AGN) \citep{Benson2003, Croton2006, Somerville2008, benson_galaxy_2010}.

Efforts to properly model these baryon flows on galactic scales typically rely on three broad categories of techniques. The most direct way involves numerical hydrodynamic simulations that explicitly track gravity and gas dynamics for particles or grid cells representing baryons and dark matter. Despite the ability to make detailed predictions, an immediate downside to this technique is that it is quite computationally expensive and, except possibly for the smallest galaxies, does not possess the necessary resolution to trace individual SNe remnants or properly track the evolution of diffuse, multiphase galactic winds. These simulations must then rely on simplified physical subgrid models to follow the physics on scales that are not resolved -- these subgrid models are generally calibrated to reproduce observed galaxy scaling relations or related observables \citep[e.g.][and the reviews by \citealt{naab_theoretical_2017} and \citealt{vogelsberger_cosmological_2019}]{2006MNRAS.373.1265O, vogelsberger_properties_2014, Crain2015}. A less costly alternative is semi-analytic modeling (SAMs) \citep[e.g.][and the reviews by \citealt{benson_galaxy_2010} and \citealt{somerville_physical_2015}]{white_core_1978, white_galaxy_1991, kauffmann93, somerville1999, cole00, baugh_primer_2006, Somerville2008}. Rather than simulating the hydrodynamics directly, SAMs track numerous interlocking physical processes in the build-up of galaxies based on the underlying dark matter merger tree using parameterized recipes. This method is relatively fast and can efficiently explore different parameter choices, but it too has its drawbacks, as SAMs can quickly become quite complex and rely on a host of free parameters that are not well-constrained. 

Despite these differences in techniques in galactic modeling, something that has been common amongst them is their reliance on strong stellar feedback to reduce the galactic baryon fraction and the overall stellar mass. Large cosmological simulations that lack the spatial resolution to model wind generation generally require a significant ejection of mass out of the galaxy in order to match observed galaxy statistics with a steep dependence on halo mass or circular velocity in order to match the stellar-to-halo mass halo relation and the low-mass end slope of the stellar mass function \citep{somerville_physical_2015}. For example, large-scale cosmological simulations such as Illustris-TNG \citep{2018MNRAS.473.4077P} or SIMBA \citep{2019MNRAS.486.2827D} explicitly impose high-mass loaded winds (although there are a few exceptions, such as EAGLE \citep{EAGLE} which injects high specific energy feedback instead). High mass outflow rates also appear in higher resolution "zoom-in" cosmological simulations such as FIRE, which aim to produce galactic winds self-consistently \citep{2015_muratov, Pandya_fire2}. Mass loadings can be even higher in semi-analytic models (10-100 times greater than the FIRE estimates) in order to compensate for the similarly larger gas accretion rate into low-mass galaxies \citep{Pandya2020_SMAUG}. 

However, this ``traditional" approach to achieving low stellar-to-halo mass fractions, by assuming that galaxies eject a large fraction of the gas which enters the disc, has come into growing tension from two directions: (i) observations of galactic winds from low-mass galaxies, and (ii) high-resolution simulations of self-consistently driven winds. 

Observationally derived mass-loading estimates for star-forming galaxies with stellar masses from $10^7 < M_{\star}/M_{\odot} < 10^{11}$ using UV absorption line measurements from the Cosmic Origins Spectrograph on board the \textit{Hubble Space Telescope} (HST-COS) find $\eta_M$ values in the range of $1-10$ with a negative correlation with stellar mass \citep{Chisholm2017}. Recent deep H$\alpha$ observations from \cite{McQuinn2019} of the low-surface brightness features of galactic winds from  a sample of 12 low-mass galaxies derive mass-loadings on the order of $0.2-7$. They find that the majority of the ejected wind from the dwarfs remain in the halo and, consistent with \cite{Martin1999}, find a much weaker dependence on circular velocity or stellar mass than what is generally adopted in cosmological simulations. Another study from \cite{Marasco2022b} using H$\alpha$ kinematics find even smaller estimates for mass-loading factors for gas that escapes the halo, reporting values as small as $\log_{\rm 10} \eta_M \sim -2$ for systems with stellar mass from $10^7 < M_{\star}/M_{\odot} < 10^{10}$. The results from \cite{Marasco2022b} appear in line with similarly low outflow rate estimates from \cite{Concas2022} at higher redshift ($1.2<z<2.6$) using stacked emission line data from the KLEVER survey.

On the simulation side, another approach to modeling galactic winds is via ``small-scale" simulations that resolve patches of the ISM down to the $\sim$pc scale and can resolve the Sedov-Taylor phase of expanding SNe remnants and their interaction with the ISM, which is crucial to resolving the multiphase nature of galactic winds \citep[e.g.,][]{SILCC2016, LiBryan2017, Kim2020}. These simulations can track outflows of mass and energy of each phase of the wind, which reach different heights from the disk as a function of gas and star formation rate densities. The consensus picture emerging from these ISM patch simulations is that the hot phase (T $\sim$ $10^6$ K) carries most of the energy, while the cold phase (T $\sim$ 10$^4$ K) carries most of the mass \citep{Fielding2018, Kim2020, li_simple_2020}. Since outflows measured in these simulations are much closer to the disk and represent a small patch of the ISM, they cannot be directly compared to observations. Mass loadings in the cool phase measured at the disk scale height can be as high as $1-100$ times the star formation rate and decreases with star formation density, but most of this cool gas is moving at low velocities, meaning that the mass loading of cooler outflows declines precipitously with increasing height from the galaxy, as cool gas that cannot escape the galaxy's potential falls back to the disc. The energy outflows in the hot phase of the wind, possessing a specific energy an order of magnitude larger than the cold phase, carry about $\sim 10\%$ of the energy ejected from SNe, enough energy to break out from the ISM and reach greater heights from the galaxy, often with Bernoulli velocities greater than the galaxy escape velocity \citep{Kim2017, Kim2020}. Lastly, there are high resolution simulations of dwarf galaxies which also find lower mass loaded winds beyond the virial radius than what is adopted in cosmological simulations \citep{Hu2019}.

These hot winds may not eject much mass out of the galaxy, but the energy they contain may contribute to a \textit{preventative} aspect of SNe feedback. Instead of only an ejective mode, energy flux from galactic winds can heat up circumgalactic material, preventing further radiative cooling in the gas and reducing the ambient density of the CGM by lifting gas to greater altitudes through gradual heating \citep{Voit2017}. This heating from a central feedback source, either from SNe or a central black hole in more massive galaxies, has been proposed as a means to regulate subsequent star formation by not only removing baryons from the CGM, preventing them from accreting onto the galaxy in the first place, but also by keeping gas in the CGM in a quasi-equilibrium state where precipitation of cold gas clouds via thermal instabilities out of a hot gaseous halo and accretion onto the disk is limited by the ratio between the cooling time $t_{\rm cool}$ and the free-fall time $t_{\rm ff}$ \citep{McCourt2012, Sharma2012b, Sharma2012a, Voit2015b, Voit2015, Voit2017}. These SN-driven winds that escape from halos (or lifted CGM gas through heating) may also shock-heat gas beyond the CGM and thereby prevent cosmic accretion \citep{Pandya2020_SMAUG}. The combination of mounting observations favoring lower mass outflows than what is required in cosmological simulations and the emerging view from small-scale simulations emphasizing the multiphase nature of galactic winds may call for a rethinking of how galaxies regulate their growth, and whether that regulation is primarily ejective or preventative in nature. 

In this paper we examine this question using a simplified method that is distinct from numerical simulations and even more stripped down than traditional SAMs for modeling the process of galaxy formation. These models have come to be known as ``gas-regulator" or ``bathtub" models \citep[e.g.,][]{bouche_impact_2010, lilly_gas_2013, dekel_toy_2013}. Explored extensively in prior works \citep{dave_analytic_2011,birrer_simple_2014, peng_haloes_2014, dekel_analytic_2014, Mitra2015, furlanetto_quasi-equilibrium_2020, furlanetto_bursty_2021}, these regulator models use simple (but physically-based) analytic arguments and approximations to model gas flows and the conversion of gas into stars. The components of the galaxy are treated as mass reservoirs (``bathtubs") which exchange mass and metals at different rates. The base framework comes from a simple set of differential equations which describe the flows between the various reservoirs. Sometimes, but not always, an equilibrium condition is also added, which states that the inflow of mass from the intergalactic medium (IGM) into the ISM balances the sum of gas mass lost from outflows and converted into stars \citep[e.g.,][]{Finlator2008}. Usually, a galaxy is assumed to evolve without mergers, with the mass inflow of dark matter and baryons set by fits to average mass accretion histories extracted from dissipationless N-body simulations. An advantage of these models is that they are simple, making clear the physical underpinnings that connect the reservoirs, and converge to solutions that are independent of initial conditions. Although gas regulator models are not meant to replace their more sophisticated cousins like SAMs or hydrodynamical simulations, these models have proved to be quite useful at reproducing basic galaxy scaling relations, and can be implemented and improved upon in simulations or SAMs in the future (Pandya et al., in prep). 

What has not been extensively explored in these regulator models is the role of the CGM in regulating galaxy formation and the contribution of energy-loaded preventative feedback. Although \cite{dave_analytic_2011}, for example, included a preventative feedback parameter, which quantifies the amount of baryonic inflow from the IGM prevented from reaching the ISM and thus remaining in the gaseous halo, the broad properties of cooling and heating in the CGM as its own reservoir are not explicitly tracked, leaving the CGM omitted on a phenomenological level. SAM models, which do explicitly include a halo gas reservoir, often assume that the thermal properties of the CGM are in equilibrium with those of their dark matter halos \citep{white_galaxy_1991}. There have been a few efforts to improve the cooling physics of halo gas in SAMs \citep{lu11, benson11, cousin2015, hou18} and generate SAM predictions for the CGM that can be compared to observables \citep{Faerman2022}, however such SAM models of galaxies and their CGMs lack an energetically self-consistent treatment of preventative feedback from galactic winds \citep{Pandya2020_SMAUG}.

While many basic properties of the CGM of the Milky Way and external galaxies are still not well understood, observations have converged onto a general picture of the CGM as an extended, multiphase reservoir composed of a hot phase ($T \sim 10^6$ K) directly measurable from X-ray emission, a warm-hot phase ($T \sim 10^{5-6}$ K) visible in UV absorption lines, a cold neutral and low-ion phase ($T \sim 10^{4}$ K), and a significant contribution of metals and dust \citep[for reviews see][]{putman_gaseous_2012,tumlinson_circumgalactic_2017}. The CGM contains a substantial budget of baryons, and could play a major role in regulating galaxy evolution as it supplies fuel to the galaxy for star formation, but the galaxy in return shapes the CGM, as outflows driven by feedback processes alter the mass and energy content of the CGM. Thus the properties of the CGM and galactic winds are deeply intertwined.

The aim of this work is to build on this understanding of the CGM as a regulator of galaxy evolution, and explore how different compositions of SNe-driven winds shape properties of galaxies and their CGMs. In particular, we will track not only the flows of mass and metallicity of the CGM, but also the flows of \textit{energy}, allowing for a self-consistent (albeit simplified) regulation of preventative feedback. We will show that the CGM is the natural seat of self-regulation in such models, with the key parameter shifting from the mass outflow rate to the energy outflow rate. Using our new model, we investigate whether we can match key observational scaling relations with reasonable assumptions about the energetics of galactic winds.

In the next section, we introduce the regulator model, along with the relevant equations governing mass evolution, the cooling efficiency of the CGM, followed by our treatment of preventative feedback from energy-loaded winds and metallicity evolution. We then show the time evolution of a $10^{12} M_{\odot}$ halo and test our model over a range of halo masses, comparing its output to the stellar-to-halo mass relation and the ISM-to-stellar mass ratio, and its implications for the circumgalactic gas content. We end our work with a discussion of its implications for self-regulation in galaxies and feedback, the connection to CGM-precipitation models, and a comparison to other regulator models. We then summarize our the conclusions.
\begin{figure*}[ht!]
    \centering
  \includegraphics[width=0.6\linewidth]{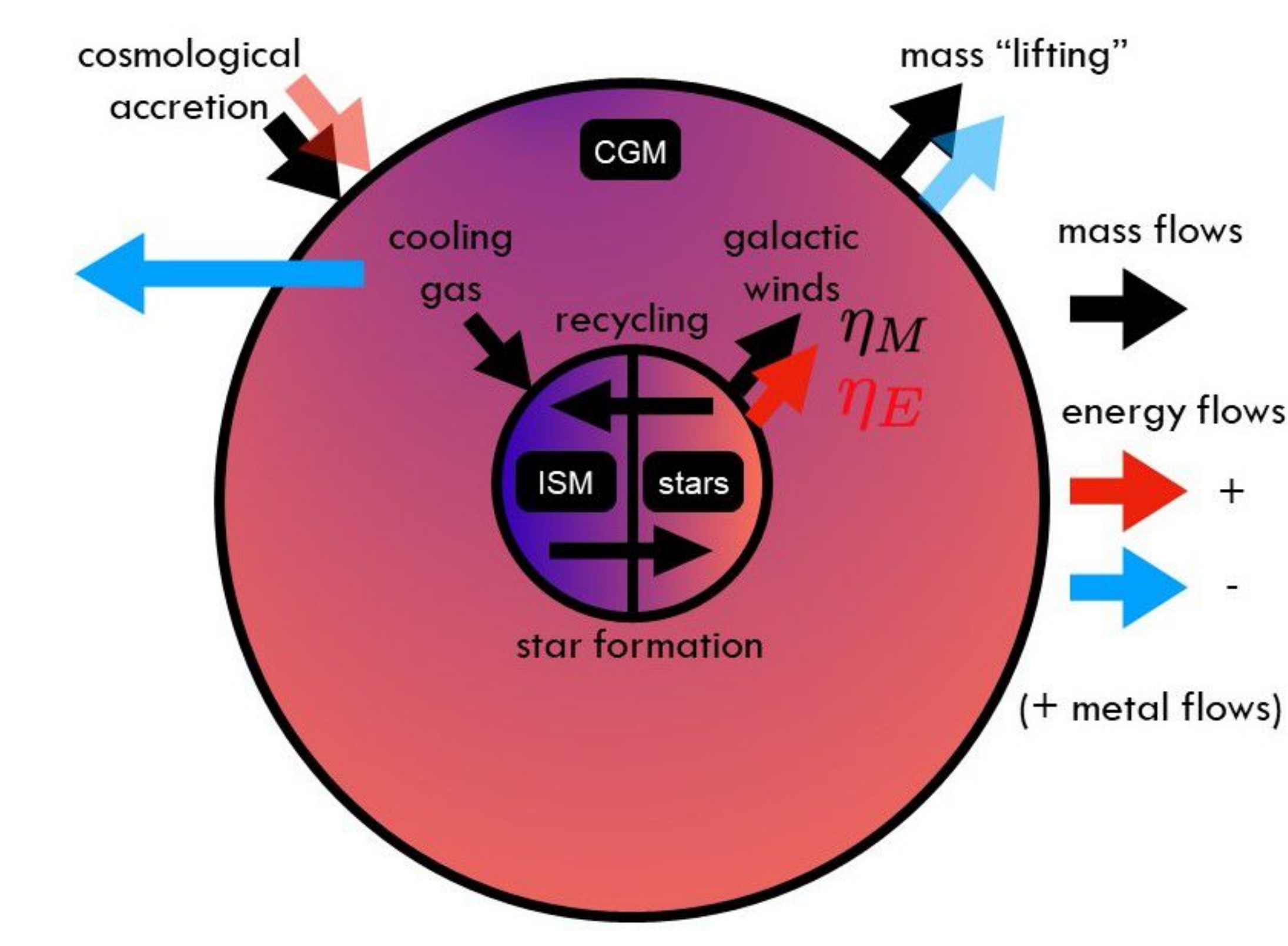}
  \caption{Schematic of 1D-regulator model. The baryonic components of the galaxy (the ISM, stellar mass, and the CGM) are distinct reservoirs, which exchange mass, energy, and metals. The black arrows represent the mass (and metal) flows, while the red/blue arrows symbolize the positive/negative flows of energy among the various components.}
  \label{fig: model}
\end{figure*}

\section{Gas-Regulator Model} \label{sec:method}
In this section, we introduce our 1D-regulator model, where the different components of the galaxy are each treated as reservoirs that exchange baryons at different rates. We consider a galaxy composed of six reservoirs (see Figure~\ref{fig: model}): a dark matter halo ($M_{\rm halo}$), the stellar mass ($M_{\star}$), the interstellar medium ($M_{\rm ISM}$), the circumgalactic medium ($M_{\rm CGM}$), the energy of the CGM ($E_{\rm CGM}$), and the mass in metals of the CGM ($M_{\rm Z, CGM}$). Note that, because we are focused on the CGM, we do not track the metallicity of the ISM or the stars, although this would be straightforward to do (but would introduce additional free parameters that do not directly impact the evolution of the CGM).

We write down a simple set of ordinary differential equations (ODEs) for the evolution of these quantities based on some simple physical reasoning (including some free parameters), and solve for their evolution using a Runge Kutta solver (using \texttt{solve\_ivp} from the \textit{scipy} python package). After introducing the basic equations that define the evolution of each of the mass reservoirs, we describe our treatment for the cooling rate for CGM gas and accretion, the lifting of CGM gas out of halo through heating from energetic winds, and the metallicity enrichment of the CGM from SNe-driven outflows. 

\subsection{Structure of the CGM}

We begin by describing our simple model for the structure of the CGM. Our goal here is to write down a model which retains the basic properties but is analytic or nearly so. We make two simplifying assumptions -- first, we take the halo to be isothermal, at temperature $T_{\rm CGM}$, which is close to but not necessarily equal to the virial temperature $T_{\rm vir}$. The temperature we assume for the CGM is equal to $T_{\rm CGM} = (\mu / k) e_{\rm CGM}$, where k is the Boltzmann constant and $\mu$ is the mean molecular weight, which we assume $\mu = 0.6$ times the mass of a proton. 
In fact, rather than the temperature, we generally use the halo specific energy $e_{\rm CGM} = E_{\rm CGM}/M_{\rm CGM}$, where $E_{\rm CGM}$ and $M_{\rm CGM}$ are the thermal energy and mass of CGM, respectively.  

In addition, we assume the density profile of the CGM follows a power-law,
\begin{align}
    \rho (r) = \rho_0 \Big( \frac{r}{r_0} \Big)^{-\alpha}, 
    \label{eq:density}
\end{align}
where $\alpha = 1.4$ so that the entropy profile goes as $K = T/\rho^{2/3} \sim r^{2/3}$ \citep{Nelson2016, Fielding2020, Esmerian2021}. As we will show, the basic behavior of the model is insensitive to reasonable variations in $\alpha$. We note that we do not explicitly enforce hydrostatic equilibrium, nor is it expected due to the presence of heating and cooling flows \citep{Stern2019, Stern2020}; however, for $T_{\rm CGM} \approx T_{\rm vir}$, this profile is close to hydrostatic equilibrium for typical dark matter halo profiles.

These simple forms allow us to write down expressions for the mass of the CGM, which we take to range from $r_0 = 0.1 r_{\rm vir}$ to $r_{\rm vir}$:
\begin{equation}
    \begin{aligned}
    M_{\rm CGM} & = \int_{r_0}^{r_{\rm vir}} \rho_0 \left(\frac{r}{r_0}\right)^{-\alpha} 4 \pi r^2 dr \\
    & = 4 \pi \rho_0 r_0^3 \left( \frac{(r_{\rm vir}/r_0)^{3-\alpha} - 1}{3-\alpha} \right)
    \end{aligned}
\end{equation}
and similarly for the net radiative cooling rate of the CGM gas:
\begin{equation} \label{eq:e_cool}
    \begin{aligned}
    \dot{E}_{\rm CGM, cool} & = \int_{r_0}^{r_{\rm vir}} \left( \frac{\rho}{\mu} \right)^2 \Lambda(T_{\rm CGM},Z,z) 4 \pi r^2 dr \\ 
    & = 4 \pi \rho_0^2 r_0^3 \left(\frac{\Lambda(T_{\rm CGM},Z,z)}{\mu^2} \right) \left(\frac{(r_{\rm vir}/r_0)^{3-2\alpha} - 1}{3-2\alpha} \right)
    \end{aligned}
\end{equation}
$\Lambda(T_{\rm CGM},Z,z)$ is the radiative cooling rate, which we take from \cite{Wiersma} and depends on metallicity and redshift (due to a time-varying metagalactic radiation field), while the mean molecular weight, $\mu$, equals 0.6 times the mass of a proton. We note that the analytic forms fail when $\alpha=3$ or $3/2$ (but the solution is well-behaved, with the terms in parenthesis becoming simply $r_{\rm vir}/r_0$).


\subsection{Mass Evolution} \label{sec:mass}

This section describes the evolutionary equations for the various mass reservoirs (halo, CGM, ISM, stellar) before turning to the CGM energy reservoir in the next section.

\subsubsection{Halo Mass}

We start with the overall evolution of the dark matter halo $M_{\rm halo}$, which we assume to grow smoothly and monotonically with time, as given by a fit to numerical simulations from \citet{dekel_cold_2009}:

\begin{equation}
    \dot{M}_{\rm halo} = 0.47 M_{\rm halo} \Big(\frac{M_{\rm halo}}{10^{12} M_{\odot}} \Big)^{0.15}  \Big(\frac{1+z}{3} \Big)^{2.25} \text{Gyr}^{-1}
    \label{eq:halo_inflow}
\end{equation}

Since halos of the same mass can have different formation histories, it is important to note that eq.\ref{eq:halo_inflow} defines the average halo growth rate for halos of a given halo mass. 

\subsubsection{CGM Mass}

We next turn to the evolution of the CGM mass component. Baryons enter the halo along with dark matter ($\dot{M}_{\rm CGM,in}$), flowing first into the CGM before cooling and accreting into the ISM ($\dot{M}_{\rm CGM,cool}$). In addition, star formation can directly eject ISM gas into the CGM ($\dot{M}_{\rm ISM,wind}$) as well as providing energy which can heat the CGM, causing it to expand and ``lift" gas out of the hot CGM and unbind it from the halo ($\dot{M}_{\rm CGM,out}$). We write this schematically as

\begin{equation}
\dot{M}_{\rm CGM} = \dot{M}_{\rm CGM,in} - \dot{M}_{\rm CGM,cool} + \dot{M}_{\rm ISM,wind}  - \dot{M}_{\rm CGM,out},
\label{eq:m-cgm}
\end{equation}

Each of these terms has relatively simple expressions and we discuss each in turn. To provide the first term, we assume that mass infalls from the IGM along with the dark matter:
\begin{equation}
\dot{M}_{\rm CGM,in} = f_b f_{\rm prevent} \dot{M}_{\rm halo}
\label{eq:m_in}
\end{equation}
where $f_b$ is the cosmic baryon fraction and $f_{\rm prevent}$ is a halo-level preventive infall factor that we describe below, after discussing the CGM energy budget. Note that this differs somewhat from preventive feedback in traditional bathtub models \citep[e.g.,][]{dave_analytic_2011} in that $f_{\rm prevent}$ is related to the accretion of gas from the IGM into the CGM, not directly to the ISM. In particular, our CGM energy model (described next) aims to explicitly calculate the traditional CGM preventive factor at the halo scale, in addition to self-consistently accounting for the prevention of accretion to the ISM by extending the cooling timescale from the CGM.

The second term accounts for radiative energy losses of the CGM that allow some of the gas to cool to low temperatures ($10^4$ K or lower) and accrete onto the ISM. For this, we write simply, 
\begin{equation}
    \dot{M}_{\rm CGM,cool} =  M_{\rm CGM}/ t_{\rm cool,eff},
    \label{eq:m_cool}
\end{equation}
where we define the effective cooling time $t_{\rm cool,eff}$ as the sum of the cooling time and the free fall at the virial radius, the second term accounting for the time the gas takes to accrete from the CGM into the ISM (this second term only becomes important at high redshift when the cooling times are short and the CGM fails to self-regulate):
\begin{equation}
t_{\rm cool,eff} =  \frac{E_{\rm CGM}}{\dot{E}_{\rm CGM,cool}} + t_{\rm ff}.
\end{equation}

The third term of Eq.~\ref{eq:m-cgm} accounts for mass ejected via a galactic wind due to star formation, which we parameterize with the usual mass loading factor $\eta_M$:
\begin{equation}
\dot{M}_{\rm ISM,wind} = \eta_M \dot{M}_{\rm SFR}.
\end{equation}
Mass ejection (generally to a long-lived reservoir) is the primary mode of self-regulation in classical regulator models, but in our model, mass ejected into the CGM can cool and accrete back to the ISM, as we will show. Instead, mass is lifted out of the CGM due to heating (from star formation and other energy sources).  Heating of the CGM can cause overpressurization (i.e. $T_{\rm CGM} > T_{\rm vir}$) which lifts gas out of the halo into the IGM, where we assume it is unbound from the halo and doesn't return.\footnote{Gas that is uplifted out of the halo may cool and return to the halo on some timescale. The inclusion of this returning gas would require greater suppression of infall (smaller $f_{\rm prevent}$) than we estimate in this work.} We will shortly lay out the energy equation for the CGM -- one term in that equation (eq.~\ref{eq:e_out}) is $\dot{E}_{\rm CGM,out}$, the rate at which energy flows out of the halo due to this overpressurization. We assume that this energetic outflow brings with it an equivalent amount of mass such that the outflowing gas has the same specific energy as the halo gas:
\begin{equation}
    \dot{M}_{\rm CGM,out} = \dot{E}_{\rm CGM,out} / (E_{\rm CGM}/M_{\rm CGM})
    \label{eq:m_out}
\end{equation}

\subsubsection{ISM Mass} 

The ISM is fed from the CGM and loses mass to star formation and SN-driven winds. 
We express the evolution of the gas mass of the galaxy in the form, 
\begin{align}
    \dot{M}_{\rm ISM} = - \dot{M}_{\rm SFR} (1+\eta_M-f_{\rm rec}) + \dot{M}_{\rm CGM,cool},
\end{align}
where the first term describes the mass lost from the ISM due to star formation and mass ejection, and the mass returned to the ISM as stars recycle a fraction of their mass back to the ISM. We fix this instantaneous stellar recycling fraction to be $f_{\rm rec}=0.4$ \citep{kroupa01}. 
The mass-loading factor $\eta_M$ quantifies how efficient star formation is at removing gas from the ISM. The second (positive) term represents gas accretion into the ISM from the CGM. The expression for $\dot{M}_{\rm CGM,cool}$ is given in Eq.~\ref{eq:m_cool}. 
\subsubsection{Stellar Mass} \label{sec: star_m}
Considering next the stellar mass reservoir, the star formation rate, $\dot{M}_{\rm SFR}$, is dictated by the gas mass of the ISM, $M_{\rm ISM}$, and the timescale over which that gas mass is converted into stars:
\begin{align}
    \dot{M}_{\rm SFR} = \frac{M_{\rm ISM}}{t_{\rm dep}},
\end{align}
where $t_{\rm dep}$ is the depletion time. For simplicity, we do not distinguish between atomic and molecular gas and assume that all ISM gas participates in star formation. We use an estimate of the depletion time as a function of stellar mass derived from observations of the cold gas mass and star formation rates of low surface brightness and star-forming galaxies from \cite{McGaugh_mainsequence_2017}. We assume a redshift dependence of $(1+z)^{-3/2}$, similar to the Hubble timescale adopted in many previous regulator models, resulting in the depletion time we use for our model:\footnote{Since the depletion times in low mass systems is not well constrained observationally, we assume that for systems with $M_{\rm star} < 5 \times 10^{7} M_{\rm \odot}$, below which the data become sparse, that the depletion time is constant and only involves with redshift and not with stellar mass.}
\begin{equation}
    t_{\rm dep} = 10^{4.92} \Bigg(\frac{M_\star}{M_{\odot}} \Bigg)^{-0.37} (1+z)^{-3/2}.
\label{eq: tdep}
\end{equation}
We can express the full evolution of the stellar mass of the galaxy: 
\begin{align}
    \dot{M}_{\star} = (1-f_{\rm rec}) \frac{M_{\rm ISM}}{t_{\rm dep}},
\end{align}
when we consider the star formation rate and the fractional loss of stellar mass from mass recycled back to the ISM.

\subsection{CGM Energy Evolution}

We turn next to the new feature of our model, an equation for the thermal energy of the CGM (we neglect here the kinetic energy component, leaving that for future work in Pandya et al., in prep). As before, we account for all of the energy gain and loss terms for the evolution of $E_{\rm CGM}$:
\begin{equation}
\dot{E}_{\rm CGM} = \dot{E}_{\rm CGM,in} - \dot{E}_{\rm CGM,cool} + \dot{E}_{\rm ISM,wind}  - \dot{E}_{\rm CGM,out},
\label{eq:e-cgm}
\end{equation}
The first term represents the energy of infalling gas (Eq.~\ref{eq:m_in}), which we assume is associated with the specific energy of the halo:
\begin{equation}
\dot{E}_{\rm CGM, in} = \left( \frac{kT_{\rm vir}}{\mu} \right) \dot{M}_{\rm CGM,in}
\end{equation}

The second term of the energy evolution equation, representing radiative losses is given in Eq.~(\ref{eq:e_cool}). The third term is positive and comes from the energy associated with SN-powered galactic winds
\begin{align}
    \dot{E}_{\rm CGM,wind} = \eta_E \dot{M}_{\rm SFR} \left( \frac{10^{51} \text{erg}}{100 M_{\odot}} \right)
    \label{eq:e_wind}
\end{align}
where we have parameterized the amount of SN-powered energy that escapes out of the ISM as $\eta_E$. As we will show, this is a key parameter that drives our CGM-regulated model.

Finally, the fourth term represents the energy that goes into lifting gas out of the CGM, unbinding it. We assume this only occurs if the CGM is overpressurized such that its specific energy exceeds the virial value. When this is true, we compute the outflow rate assuming that gas flows out at the hot gas sound speed $c_s$:
\begin{equation}
    \dot{E}_{\rm CGM,out} = {\rm max} \left( E_{\rm CGM} - k T_{\rm vir} M_{\rm CGM}/\mu , 0 \right) \left( \frac{c_s}{R_{\rm vir}} \right)
    \label{eq:e_out}
\end{equation}
This outflow of hot gas also drives a mass flux (see eq.~\ref{eq:m_out}). The hot gas sound speed is the usual $c_s = (5 k T_{\rm CGM}/3\mu)^{1/2}$.

The inclusion of this mass and energy ``lifting" term to the CGM describes the contribution from preventative feedback. By lifting a fraction of the CGM beyond the virial radius and (we assume) out of the gravitational clutches of the dark halo, feedback works to reduce the density of the CGM. This lengthens the radiative cooling timescale, and thus regulates the rate of accretion out of the circumgalactic medium onto the galaxy, and any subsequent star formation. The details of this ``lifting" process are clearly complicated and the simple relation in Eq.(\ref{eq:e_out}) is an approximation; however, small uncertainties in the proportionality can be accommodated by changing the effective energy-loading factor.

One of the last elements of the model is to define the halo-level preventive inflow factor $f_{\rm prevent}$ used in Eq.~\ref{eq:m_in}. The idea here is that energy flowing out beyond the virial radius will heat IGM gas and prevent it from falling in. This effect has previously been found to be operating in the FIRE simulations \citep{Pandya2020_SMAUG}. We take a deliberately simple form for this, defining it to be proportional to the ratio of the inflowing to outflowing energy. Although there is no strong physical justification for this particular form, it seems reasonable that this ratio of these energies should be a good measure of the importance of energetic outflows. In particular, we adopt:\footnote{Note that here $\dot{E}_{\rm CGM,in}$ is defined as the energy of inflowing gas from the IGM at the full cosmic baryon fraction.}
\begin{equation}
    f_{\rm prevent} = {\rm min} \left(\alpha_{\rm prevent} \frac{\dot{E}_{\rm CGM,in}}{ \dot{E}_{\rm CGM,out}}, 1 \right)
\end{equation}
where we have defined a free parameter which we take to be $\alpha_{\rm prevent} = 2$, resulting in $f_{\rm prevent}$ values consistent with \citet{Pandya2020_SMAUG}.

\subsection{Metallicity of the CGM} \label{sec: 2.4}
The significant share of metals observed in the CGM of galaxies is incontrovertible evidence for the presence of metals in galactic outflows \citep{prochaska_cos-halos_2017}. Metals in galactic winds arise, in part by entrainment of previously enriched ISM gas, and in part to the fact that SNe enrich the gas as it simultaneously injects energy and momentum. We focus on the second source and parameterize the fraction of metals produced by SNe that wind up in outflows using the \textit{metal loading factor}, defined as $\eta_Z = \frac{\dot{M}_{Z, out, SN}}{\dot{M}_{Z, SN}}$. This definition of the metallicity-loading factor means that if $\eta_Z = 1$, then all of the metals produced in SNe leave the galaxy in outflows. We remark that we are assuming that this factor is independent of the metallicity of the ISM, which is clearly an oversimplification; however in simulations which involve a high-energy loading factor, much of the metals come directly from the SN ejecta, rather than via entrainment of the ISM, something that will have interesting implications for the detailed CGM abundances, but is beyond the scope of the current work. 

If we include the metal mass of the CGM ($M_{\rm Z, CGM}$) as its own reservoir, its evolution takes the form: 
\begin{align}
    \begin{gathered}
    \dot{M}_{\rm Z, CGM} = \eta_Z y_{\rm SN} \dot{M}_{\rm SFR} + Z_{\rm IGM} \dot{M}_{\rm CGM,in} \\ - f_{\rm Z, CGM} (\dot{M}_{\rm CGM,cool} + \dot{M}_{\rm CGM,out}).
    \end{gathered}
\end{align}
The first term captures the enrichment of metals from SNe-outflows, where the metal mass enrichment yield produced from one supernova per 100 $M_{\odot}$ is approximately equivalent to $y_{\rm SN} \sim 0.02$. The second term is the accretion of metals from the IGM. We assume a constant metallicity for the IGM inflow of $Z_{\rm IGM} = 0.01 Z_{\odot}$, where the IGM metallicity is (somewhat arbitrarily) assumed to be 1$\%$ of the solar metallicity $Z_{\odot}=0.0134$ \citep{2009_asplund_sun}. This is different from the usual assumption that the IGM metallicity is "pristine" and has only been enriched by Pop III SNe. Here we imagine that some of the cosmic accretion has been further pre-enriched either by the galaxy's own outflows or from galactic neighbors. 

Lastly, the third term captures the metals lost to accretion into the ISM and those lifted out of the halo, with $f_{\rm Z, CGM} = M_{\rm Z,CGM}/M_{\rm CGM}$ representing the fraction of the CGM mass in metals. 
 
\section{Results} \label{sec:results}

Now that we have described our basic model, we explore it in detail for a single halo, before turning to the scaling relations that it predicts as the halo mass is varied. 

\begin{figure}
	\centering
  	\includegraphics[width=0.87\linewidth]{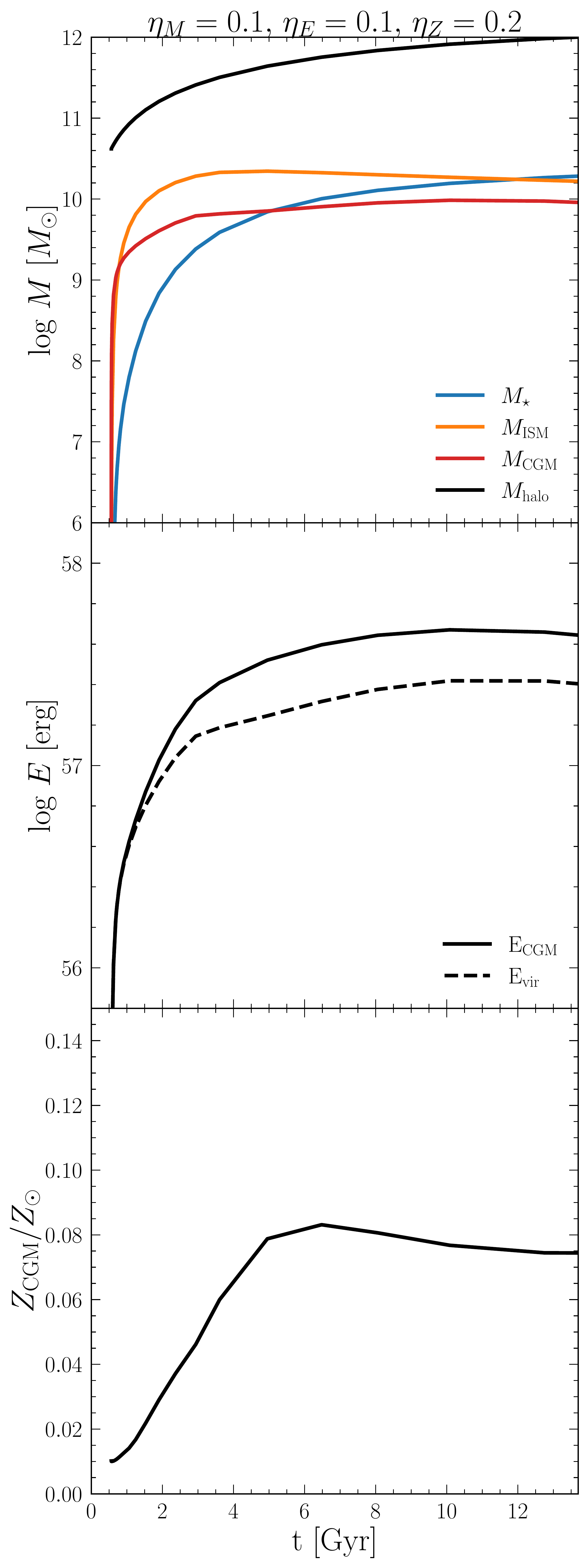}
    \caption{Time evolution of the mass of each reservoir in the regulator model (\textbf{upper}), the total energy of the CGM (solid) compared to its virial binding energy (dashed) (\textbf{middle}), and the metallicity of gas in the CGM with respect to the solar metallicity (\textbf{lower}) for a $10^{12} M_{\odot}$ mass halo from $z \simeq 6$ to $z \simeq 0$. The mass, energy, and metallicity loadings are equal to $\eta_M = 0.1$, $\eta_E = 0.1$, and $\eta_Z = 0.2$ respectively.}
    \label{fig: halo12}
\end{figure}

\subsection{Evolution for a $10^{12}$ M$_{\odot}$ Halo} \label{sec: 3.1}
To get a sense of whether the regulator model can reproduce galaxy properties by the present day, we run our model on a Milky Way-like galaxy. The top panel of Figure~\ref{fig: halo12} shows the time-evolution of the mass in each of our four reservoirs, for a $10^{12} M_{\odot}$ halo from a redshift of $z\simeq6$ to the present-day. The middle panel charts the evolution of the total CGM energy compared to its virial binding energy ($k T_{\rm vir} M_{\rm CGM}$). The bottom panel displays the CGM metallicity as a function of time with respect to solar metallicity. This run uses parameter values of $\eta_M = 0.1$, $\eta_E = 0.1$, and $\eta_Z = 0.2$ respectively, in line with typical values for the loading factors in MW-like galaxies seen in FIRE-2 \citep{Pandya_fire2}. 

There is rapid growth in the CGM and ISM masses early on, as pristine gas from the IGM accretes onto the CGM before rapidly cooling and accreting onto the central galaxy. This prompts a steep initial rise in the gas mass of the galaxy, and this newly available fuel triggers star formation on a timescale of order $t_{\rm dep}$. The birth of new stars increases the stellar mass, but their subsequent death through supernovae begins to drive energy and mass outflows back into the CGM. Due to the relatively small mass-loading factor adopted here, the major contribution to the CGM is due to the energy-loading, which heats the CGM and lifts mass out of the halo, into the IGM. This loss of mass from the CGM not only serves as a counterbalance to the mass gains from early IGM accretion by limiting the mass of the CGM across a wide range of redshifts, but also regulates the density of the CGM and therefore its overall cooling efficiency. This regulation of cooling limits the flow of gas into the ISM, resulting in generally slow growth of the ISM and stellar masses. 

The model produces stellar and ISM masses on the order of $\sim 10^{10} M_{\odot}$, estimates that are slightly lower but broadly consistent with the total stellar mass of the Milky Way of $\sim 5 \times 10^{10} M_{\rm \odot}$ \citep{Bland-Hawthorn2016}. The mass estimate of the CGM is also on order of $\sim 10^{10} M_{\odot}$, lying within the observational constraints of the CGM mass of the Milky Way for reasonable choices of $\eta_M$ and $\eta_E$ \citep[e.g.,][]{salem_ram_2015, tumlinson_circumgalactic_2017}. However, we note that the total CGM mass is poorly constrained observationally.

The total energy of the CGM grows rapidly at early times from the infall of gas from the IGM. Radiative losses cause the gas to cool and accrete quickly into the ISM, where the early burst of star formation drives energy outflows back into the CGM. The energy-loading of these outflows heats the CGM, raising its specific energy and limiting its cooling efficiency. Comparing the time evolution of the total energy of the CGM against its virial energy, we find that the total energy of the CGM exceeds its virial value, resulting in a CGM that is overpressurized. This overpressurization expands the CGM, lowering its density and lifting mass and energy beyond the virial radius. This brings the energy of the CGM closer to its virial value over time. This process of heating the CGM and lifting gas out of the halo through overpressurization regulates the energy content of the CGM against radiative loses and energy gains from cosmic accretion and star formation, thus limiting the growth of the CGM energy to the present-day. 

The growth of $M_{\star}$ produces metals, and after a steep initial climb from the first SNe-driven outflows of metals, the metallicity of the CGM slowly decreases with time. The on-going star formation continues to increase the supply of metals being delivered to the CGM. Metal-rich outflows work to enhance future cooling from the CGM, while efficient energy-loading work to reduce it. The metallicity of the CGM with respect to the solar metallicity approaches $Z/Z_{\odot} \simeq 0.07$ by $z=0$, lower than existing estimates of the metallicity of the Milky Way's CGM and similar hosts \citep{prochaska_cos-halos_2017}. This suggests that either larger metallicity outflows are needed to match observations or a consequence of the fact that we don't consider the metallicity evolution of the ISM and its outflows. 

Our regulator model can produce galaxies by the present-day with properties that are in qualitative agreement with observations of the Milky Way-like galaxies. This was achieved through mass outflows out of the ISM and energy regulation of the CGM acting as a mode of preventative feedback by constraining the total baryonic content of the galaxy.

\subsection{Galaxy Scaling Relations} \label{sec: 3.2}
Next we compare the $z=0$ galaxy properties predicted from our model to galaxy scaling relations in the range $M_{\rm halo} = 10^{10} - 10^{12} M_{\odot}$. SNe are believed to drive the primary mode of feedback and star formation self-regulation for galaxies in halos below $\sim 10^{12} M_{\odot}$, so the regulator model should be compared in this range. Since the effects of AGN-driven feedback are not considered in this work, disparities at the high mass end of the considered range may be present between the data and the model. 

\begin{figure*}
	\centering
  	\includegraphics[width=1\textwidth]{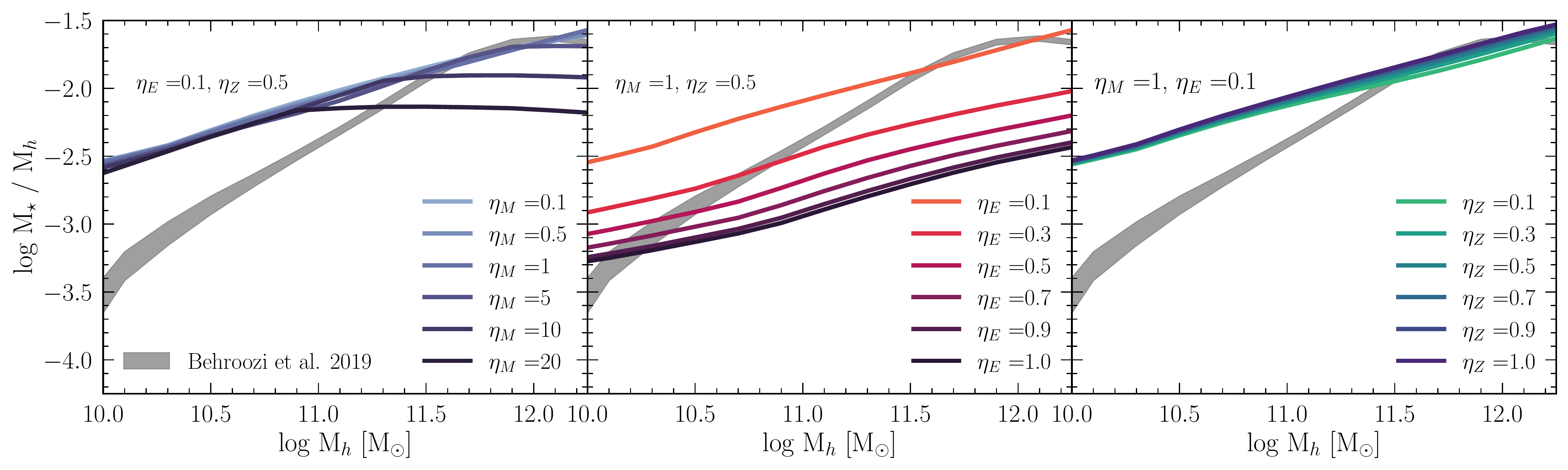}
    \caption{Log of $M_\star / M_{\rm h}$ from the regulator model as a function of halo mass, compared to the median observed stellar mass with uncertainties in gray and their derived halo mass from abundance matching from \citet{2019_behroozi}. Estimates of $M_\star / M_{\rm h}$ are presented as a function of increasing the power-law dependence of wind mass-loading $\eta_M$ (\textbf{left}), energy-loading $\eta_E$ (\textbf{middle}), and metallicity-loading $\eta_Z$ (\textbf{right}). The annotations of each plot display which parameters were kept constant in the iterations of the model on display.}
    \label{fig: MsMh}
\end{figure*}

\subsubsection{$M_\star / M_{\rm halo}$-$M_{\rm halo}$ Relation} \label{sec: 3.2.1}

We display the predicted $M_\star / M_{\rm halo} - M_{\rm halo}$ relation over this range for a set of different outflow parameters in Figure \ref{fig: MsMh}, plotted alongside the estimated observed stellar masses with uncertainties and their derived halo mass from abundance matching from \citet{2019_behroozi}. The relation rises as a power-law until peaking at a mass of $10^{12} M_{\odot}$, where the stellar mass fraction is largest, and then steadily falls for more massive halos as the quenching effects of AGN take hold. We choose an initial set of outflow parameters, setting $\eta_M = 1$, $\eta_E = 0.1$, and $\eta_Z = 0.5$, which are typical In our exploration, we modulate the strength of one outflow parameter at a time, while keeping the values of the other two set to their fiducial values.  


As the left panel demonstrates, the predicted stellar mass for low-mass halos is remarkably robust against changes in $\eta_M$, as the mass-loading factor is increased by a factor of 200; the only significant impact is to reduce the stellar mass of the highest mass halos when using very highly mass-loaded winds. This suggests that the system \textit{self-regulates} to similar galactic properties despite large changes in $\eta_M$. Self-regulation in this instance means that enhancements in the mass-loading of outflows increase the mass and density of the CGM, which increases the CGM's ability to cool and accrete gas back to the galaxy. When $\eta_E$ is held constant, highly mass-loaded outflows also lower the specific energy of the CGM. This lowers the overpressurization of the CGM and thus the flux of mass and energy leaving the halo, resulting in a larger $f_{\rm prevent}$ factor at the virial radius. Both of these processes$-$weakening the suppression of mass infall from the IGM and enhancing radiative losses of CGM gas$-$restrict the ability of mass-loaded outflows to meaningfully reduce the present-day stellar mass of galaxies on their own. The largest $\eta_M$ values explored in Figure \ref{fig: MsMh} do appear to be more effective at reducing the overall stellar mass in more massive halos, which may seem counter-intuitive. The $M_\star/M_{\rm halo}$ ratio flattens at halo masses beyond a threshold mass, which becomes lower as $\eta_M$ increases. This occurs when $\eta_M$ is large enough at a given halo mass (and given $\eta_E$) such that the outflow specific energy (and so $e_{\rm CGM}$) drops below $kT_{\rm vir} / \mu$, which shuts off all energy and mass outflows at the virial radius (Eq. \ref{eq:e_out}) and with that, any preventative feedback at the halo scale ($f_{\rm prevent} = 1$). The impact of larger values of $\eta_M$ on the stellar mass beyond that threshold are still countered by enhancing CGM density and cooling, but are no longer affected by any cosmic outflows. 

The center panel of Figure \ref{fig: MsMh} displays the influence of changing $\eta_E$. Increasing the energy-loading factor of the outflows has the most significant impact of all outflow parameters, reducing $M_{\star}$ for larger $\eta_E$. Increasing $\eta_E$ has the effect of enhancing the mass and energy lifted out of the CGM, as well as heating a larger share of IGM gas beyond $R_{\rm vir}$ and preventing its accretion. This reduces the CGM mass and density, and thus its ability to cool. Lowering the accretion rate reduces the overall star formation rate, resulting in much lower estimates of the present-day stellar mass for all halos. Since the mass lifted out of the halo is assumed to never return to the galaxy in our model, increasing $\eta_E$ has the effect of permanently removing baryons from the galactic system. It is also worthy of note that the model is able to produce stellar masses in the lowest mass galaxies approaching those of observations with $\eta_E \sim 1$. This is despite having mass-loadings below unity, which is at least an order of magnitude smaller than what is usually required to match galaxy scaling relations at the low-mass end.  Interestingly, plotting the $M_\star / M_{\rm halo} - M_{\rm halo}$ relation from the regulator model for different constant values of $\eta_E$ against the observed relation reveals that in order to reproduce the halo mass dependence of the observed relation, $\eta_E$ must depend on halo mass, favoring $\eta_E$ close to unity for the lowest mass halos and $\sim 0.1$ for Milky Way-like halos. 

As for changing $\eta_Z$ (rightmost plot), we find only a very small effect on the stellar mass of the galaxy in the halo mass range explored. Increasing the metallicity of the winds increases the overall cooling efficiency of the CGM. However, because of the tight energy regulation, this enhanced cooling does not result in a significantly larger stellar mass (the mean density of the CGM is decreased by feedback to counterbalance the enhanced metallicity, resulting in nearly the same total cooling and star formation). The cancellation is such that this parameter appears to only make a meaningful difference in the accretion rate in halos with high star formation (and more massive halos). 

The outflow parameter with the most significant influence on the present-day stellar mass is $\eta_E$, likely pointing to its prominent role in regulating the density of the CGM and limiting future gas cooling through preventative feedback. This form of preventative feedback becomes increasingly important in lower mass halos. We will parameterize and discuss this dependence between $\eta_E$ and $M_{\rm halo}$ in Section~\ref{sec: bisect}. 


\begin{figure*}
	\centering
  	\includegraphics[width=1\textwidth]{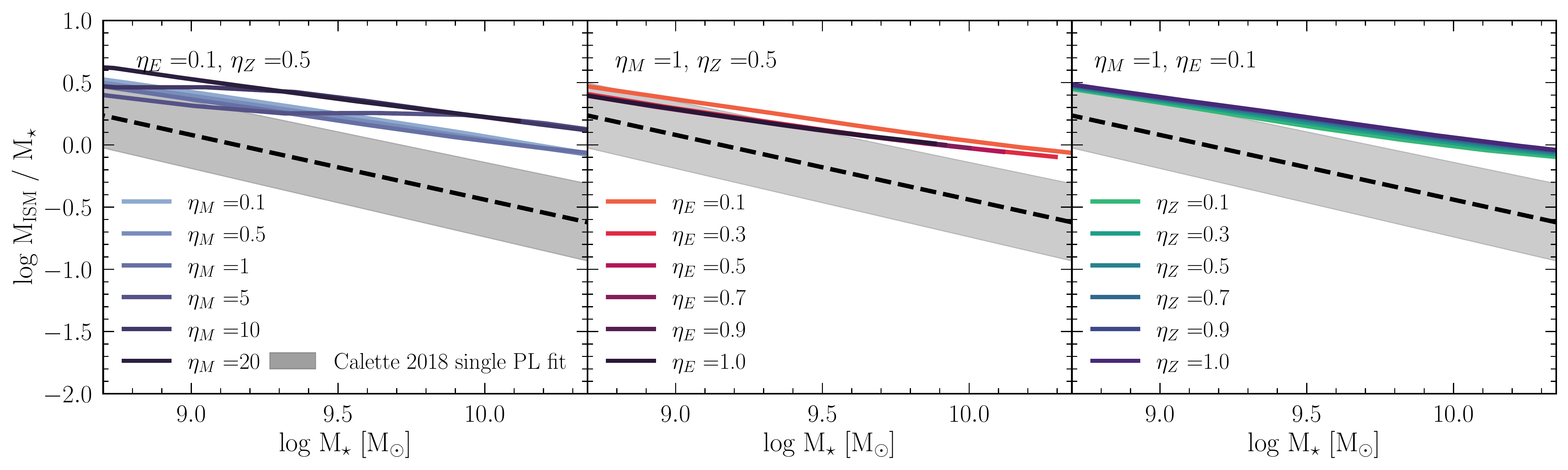}
    \caption{Log of $M_{\rm ISM} / M_\star$ from the regulator model as a function of stellar mass, compared to the best-fit single power-law fit for the total cold gas mass as a function of $M_\star$ over the range $7.3\lesssim\log{(M_\star / M_{\rm h})}\lesssim11.2$ for late-type galaxies from \citet{calette_hi-_2018}. Estimates of $M_{\rm ISM} / M_\star$ are presented as a function of increasing the power-law dependence of wind mass-loading $\eta_M$ (\textbf{left}), energy-loading $\eta_E$ (\textbf{middle}), and metallicity-loading $\eta_Z$ (\textbf{right}). The annotations of each plot display which parameters were kept constant in the iterations of the model on display.}
    \label{fig: MgMs}
\end{figure*}


\subsubsection{$M_{\rm ISM}/M_\star$-$M_\star$ Relation}  \label{sec: 3.2.2}
Now we turn to the relation between the ISM gas and the stellar mass, which encodes star formation efficiency and can be constraining on models of galaxy evolution. Figure~\ref{fig: MgMs} compares the best-fit single power-law fit for the total cold gas mass as a function of $M_\star$ over the range $7.3\lesssim\log{(M_\star / M_{\rm halo})}\lesssim11.2$ for late-type galaxies from \cite{calette_hi-_2018} to the gas-to-stellar mass relation at $z=0$ for the regulator model.

Figure \ref{fig: MgMs} shows the $M_{\rm ISM}/M_\star$-$M_\star$ relation as a function of outflow parameters. Changes in the energy (middle panel) and metallicity (right panel) composition of the outflows makes negligible difference to the proportion of gas mass to stellar mass over the parameter and halo range studied. The mass-loading factor (left panel) makes the most discernible difference in the $M_{\rm ISM}/M_\star$-$M_\star$ relation, where more efficient mass loadings somewhat increase the ratio of gas mass to stellar mass for all galaxies and leads to a flatter dependence with stellar mass. The larger mass loading lowers the specific energy of the CGM and inhibits the suppression of cosmic inflow (larger $f_{\rm prevent}$ means more gas falls in), especially in more massive halos. This results in a more massive CGM and inevitably a more massive ISM reservoir due to the prompt radiative loses of the halo gas.   

Our current treatment of the star formation produces a shallower gradient with increasing stellar mass than the observed power-law fit in \cite{calette_hi-_2018}. This means that we predict an excess of gas with increasing halo mass. Although this could stem from a sampling bias in the data where we derive our depletion time, or from uncertainties in HI or molecular gas content, the size of the mismatch seems significant. However, as we discuss in the appendix, the slope and amplitude of this relation in our model depends greatly on our choice of the depletion time. This dependence makes sense because the depletion timescale governs the rate at which gas in the ISM is converted to stars. Additional feedback from AGN may also help to ameliorate these conflicts at the higher mass end, providing another means of reducing the ISM mass and quenching star formation. As demonstrated in the appendix, we are able to vary the depletion time to match this relation and are still able to produce galaxies with stellar masses in qualitative agreement with observational constraints.


\begin{figure*}
	\centering
  	\includegraphics[width=1\textwidth]{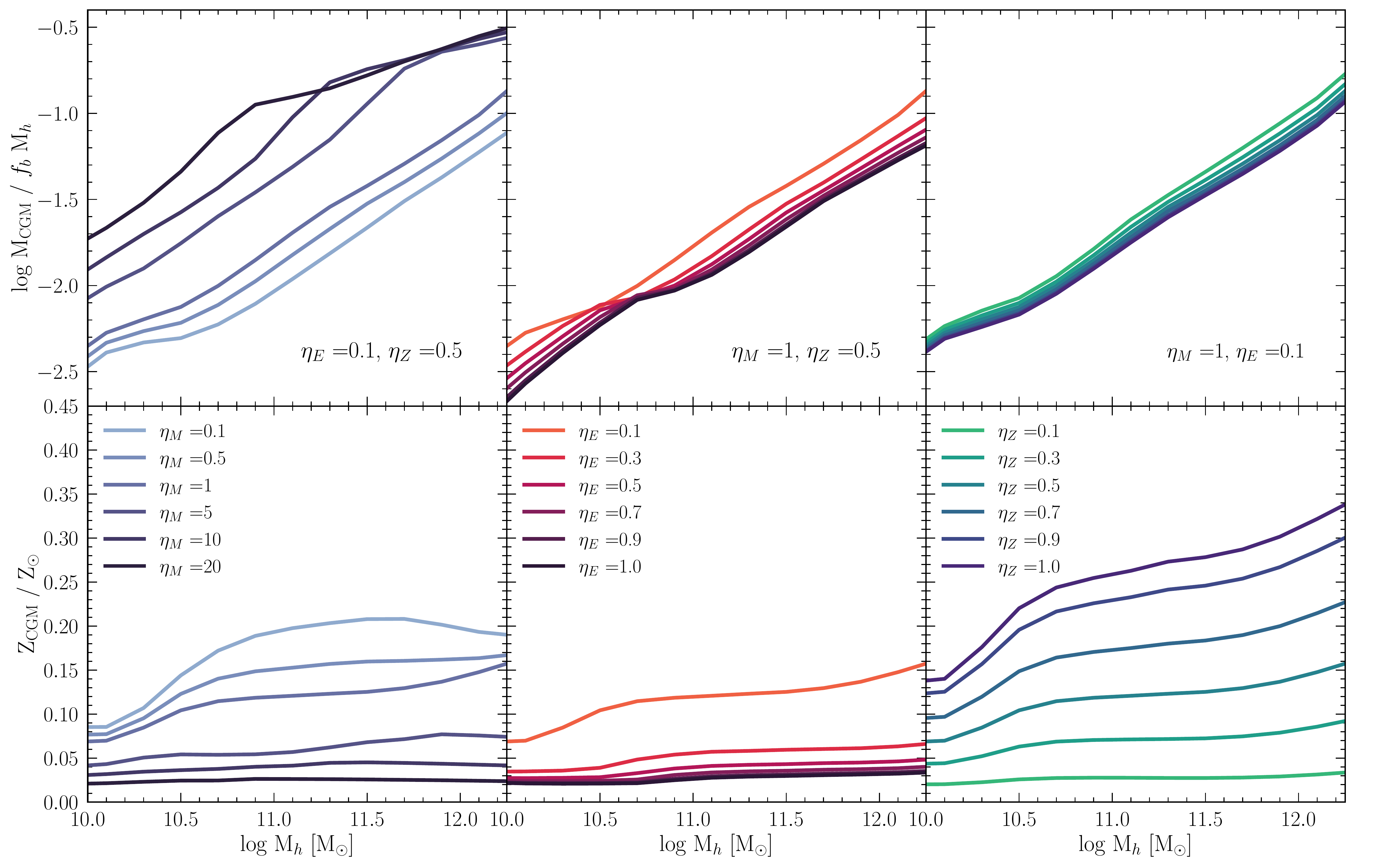}
    \caption{Estimates of $\log(M_{\rm CGM} / f_b M_{\rm h})$ and the CGM metallicity with respect to the solar metallicity from the regulator model as a function of halo mass. Estimates of $M_{\rm CGM} / M_{\rm halo}$ and $Z/Z_{\odot}$ are presented as a function of increasing the power-law dependence of wind mass-loading $\eta_M$ (\textbf{left column}), energy-loading $\eta_E$ (\textbf{middle column}), and metallicity-loading $\eta_Z$ (\textbf{right column}). The annotations of each plot display which parameters were kept constant in the iterations of the model on display on a column.}
    \label{fig: Mcgm}
\end{figure*}

\subsubsection{Implications for the CGM gas content}  \label{sec: 3.2.4}

The properties of the CGM and the properties of galactic outflows are intertwined, such that changes to the composition of galactic winds ought to greatly influence the properties of the CGM. Figure~\ref{fig: Mcgm} plots the mass (upper panel) and metallicity of the CGM (lower panel) as a function of halo mass for different wind mass-loadings. Broadly we find a that the CGM mass is mostly dependent on the mass-loading while less sensitive to the adopted energy-loading factor (in contrast to what was seen for the stellar-mass halo-mass relation), and that the metallicity drops as $\eta_M$ or $\eta_E$ increases, while it (unsurprisingly) increases with increasing $\eta_Z$. In the following paragraphs, we examine these trends more carefully.

First, we see with the leftmost panels of Figure \ref{fig: Mcgm}, that increasing $\eta_M$ has the fairly predictable effect of increasing the overall CGM mass (although not linearly with $\eta_M$). This is true for all halo masses in our range. Increasing the mass loading has the effect of reducing the overall metallicity of the CGM in our model. Since $\eta_Z$ is kept constant and the star formation rate is largely insensitive to wind mass-loading, large mass-loading dilutes the average metallicity content of the wind and thus its contribution to the CGM metallicity. However, since we don't consider the metallicity evolution of the ISM and in turn, the metallicity of the mass-loaded outflow, the full consequences for the CGM metallicity in real galaxies for increasing $\eta_M$ alone are not completely captured by our model. 

Exploring the variation of the energy-loading in the middle column, we see that more efficient energy-loading has the opposite effect, reducing the overall mass of the CGM as energetic flows from SNe heat the CGM and lift gas out of the halo. The reduction of star formation in the high $\eta_E$ scenario also stymies overall metal production in the galaxy, leaving less to be ejected into the CGM in the first place. Similar to the case of raising $\eta_M$, the lowest mass halos are quite metal-poor, producing metallicities by the present-day that are not much higher than what we assumed for the background metallicity of the IGM of $Z_{\rm IGM}/Z_{\odot} = 0.01$. This is consistent with the interpretation that dwarfs easily lose most of the metals that their stars produced in outflows to the IGM \citep{Muratov2017, Pandya_fire2}.

Increasing $\eta_Z$ has a weak impact on the CGM mass but a strong influence on the metallicity: building up the metallicity of the CGM in Milky Way halos to $\sim 35\%$ of the solar metallicity, in rough agreement with metallicity estimates from COS-Halos survey \citep{prochaska_cos-halos_2017}, but higher $\eta_Z$ still leaves the CGM metallicity around $10\%-15\%$ of the solar metallicity in the lowest mass halos of our range. When considering the mass, metal-enriched winds do have the effect of increasing the cooling efficiency in the CGM, leading to more mass lost via cooling and accretion in the galaxy. This increase in cooling efficiency is small in the low-mass halos, but becomes slightly more important for halos above $\log(M_{\rm halo}/M_{\odot}) \gtrsim 10.5$. 

When we assume $\eta_E \lesssim 0.1$ and/or $\eta_M > 1$, our CGM mass estimates for Milky-Way mass galaxies approach those of the models presented in \cite{Faerman2020, Faerman2022}, where they find MW CGM mass estimates in the range of $\sim 3-10 \times 10^{10} M_{\odot}$. These models have been shown to agree with a broad variety of observational constraints on the Milky Way CGM.  However, our mass estimates for dwarf CGMs with those fixed parameters disagree with the values found in the FIRE-2 simulations \citep{Pandya2020_SMAUG}. CGM masses at the low-mass end are sensitive to not only the mass loading, which in FIRE is at least a factor of 2 higher than our maximum assumed $\eta_M = 20$ but also the shape of the cooling curve. The lowest CGM masses are found in dwarfs where the CGM temperature approaches the peak of the cooling curve. 

Since we assume an isothermal CGM with temperature derived from the specific energy, we cannot model the variations in gas phases captured by observations, but our work does have important implications for the CGM's multiphase structure. Gas that cools and precipitates out of a hot halo onto the galaxy will be transitioning to cooler, more dense gas phases, forming structures that could resemble the high-velocity cool cloud complexes observed in the Milky Way's CGM \citep{Maller2004, Fraternali_hail, Li_Tonnesen2020}. On the hotter end, energetic outflows could heat gas from cool to warm phases (preventing its accretion) as well as heat warmer gas to super-virial temperatures before it ultimately becomes unbound from the halo \citep{das_hot_2021}.

\subsection{The Halo-Mass Dependence of Wind Energy-loading} \label{sec: bisect}
\begin{figure}[h]
	\centering
  	\includegraphics[width=0.77\linewidth]{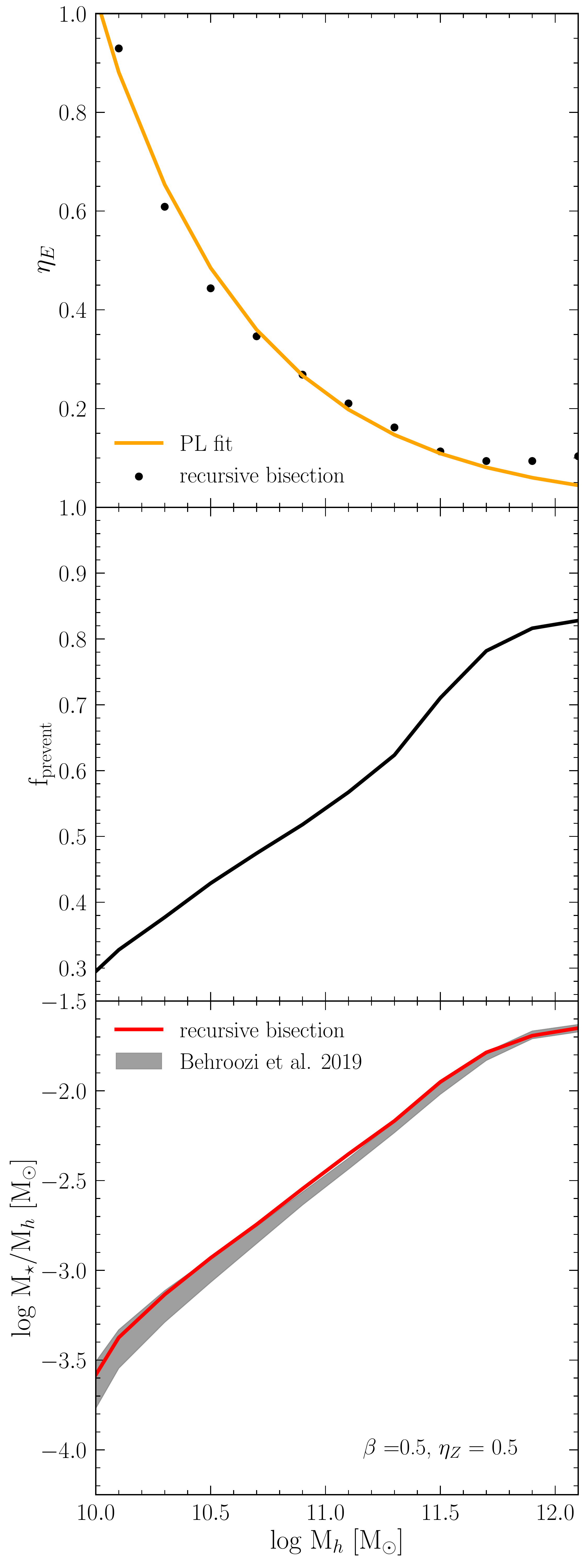}
    \caption{Energy loading $\eta_E$ as a function of halo mass from recursive bisection (points) plotted with its best-fit power-law fit (red) (\textbf{top}). The \textbf{middle} shows the preventative inflow factor $f_{\rm prevent}$ as a function of halo mass for the optimal choice of $\eta_E$. Log of $M_{\star}/M_{h}$ from the regulator model computed with $\eta_E$ from recursive bisection as a function of halo mass, compared to the median observed stellar mass with uncertainties in gray and their derived halo mass from abundance matching from \cite{2019_behroozi} (\textbf{bottom}). The annotations on the bottom plot display which parameters were kept constant.}
    \label{fig: bisec}
\end{figure}

Comparisons of the regulator model with the $M_\star / M_{\rm halo} - M_{\rm halo}$ relation in Figure \ref{fig: MsMh} indicate that $\eta_E$ must depend strongly on halo mass. To ascertain the form of this dependence, we perform a recursive bisection using \texttt{optimize.bisect} from the \textit{scipy} python package to find the optimal value of $\eta_E$ that would yield the closest agreement between the model and the $M_\star / M_{\rm halo} - M_{\rm halo}$ relation at z=0 from \cite{2019_behroozi}. 

Previous work on stellar feedback in low-mass galaxies from simulations and observations find that mass loadings likely depend on halo mass \citep[see recent review][]{2022NatAs...6..647C}. For our recursive bisection on $\eta_E$, we assume that the mass-loading factor of outflows has a simple power-law dependence with halo mass with the form
\begin{align}
    \eta_M = \Big( \frac{M_{\rm halo} (z=0)}{10^{12} M_{\odot}}\Big)^{-\beta}
    \label{eq: eta_M}
\end{align}
where $\beta$ is a power-law index that we set to $\beta = 0.5$. This value of $\beta$ translates to larger mass-loadings for lower mass halos, where $\eta_M = 10$ and $\eta_M = 1$ for halos of mass $\sim 10^{10} M_{\odot}$ and $\sim 10^{12} M_{\odot}$ respectively. We assume this scaling for $\beta$ such that our values for our mass-loading factor are comparable to those derived from observations of galactic winds in nearby dwarf and Milky Way-like galaxies \citep{Chisholm2017, McQuinn2019}. We also assume $\eta_Z = 0.5$, the same as in section \ref{sec:results}. Metal-loading may also scale with halo mass in real galaxies, but as we demonstrated in section \ref{sec: 3.2.1}, our results for the stellar-to-halo mass relation are not strongly dependent on our choices for $\eta_M$ and $\eta_Z$, and should not significantly alter our fitted value for $\eta_E$.

We plot the optimal values of $\eta_E$ from our recursive bisection to match the median stellar mass at a given halo mass at $z=0$ in the top panel of Figure~\ref{fig: bisec}, along with a best fit power law relation of $\eta_E$ from our best-fit model as a function of halo mass using \texttt{optimize.curve\_fit}. This power law relation takes the form: 
\begin{align}
    \eta_E = A \Big( \frac{M_{\rm halo}(z=0)}{10^{12} M_{\odot}}\Big)^{-\lambda},
    \label{eq: eta_E}
\end{align}
and find good agreement with fitted parameter values of $A = 0.051 \pm 0.007$ and $\lambda = 0.65 \pm 0.04$. We find that our model prefers $\eta_E \sim 1$ for the lowest mass halos and $\sim 0.1$ for MW-mass halos.

Our best-fit model also yields an estimate for the total CGM preventative inflow factor $f_{\rm prevent}$ as a function of halo mass. Displayed in the middle panel of Figure \ref{fig: bisec}, we find that mass and energy outflows from the halo suppresses cosmic inflow to well below the cosmic baryon fraction, reaching $f_{\rm prevent} \sim 0.3$ in the lowest mass halos. Preventative inflow factors of $f_{\rm prevent} \sim 0.3$ are also found in FIRE for similarly low-mass halos \citep{Pandya2020_SMAUG}, and rise to $f_{\rm prevent} \sim 1$ for Milky Way mass halos, slightly higher than the results of our model where $f_{\rm prevent} \sim 0.8$ in those halos. In the bottom panel, we show the resulting model stellar mass-halo mass.

The relation we find, in which $\eta_E$ decreases in more massive halos appears consistent with the results of FIRE-2, which also find larger energy-loadings in low-mass dwarfs relative to MW-like halos at both high and low redshifts \citep{Pandya_fire2}. This result may conflict with results from ``tall-box" simulations like TIGRESS which show a flat relationship between $\eta_E$ and star formation density and lower values of $\eta_E$ overall ($\eta_E \sim 0.1$) \cite{li_simple_2020}. These ISM patch simulations were done under Milky Way-like conditions with solar metallicities \citep{Kim2020}. Simulations that can probe the interaction between SNe and the ISM under low-metallicity conditions will be needed to better constrain $\eta_E$ in low-mass galaxies.

\section{Discussion} \label{sec: dis}

\begin{figure*}[h]
	\centering
  	\includegraphics[width=0.95\textwidth]{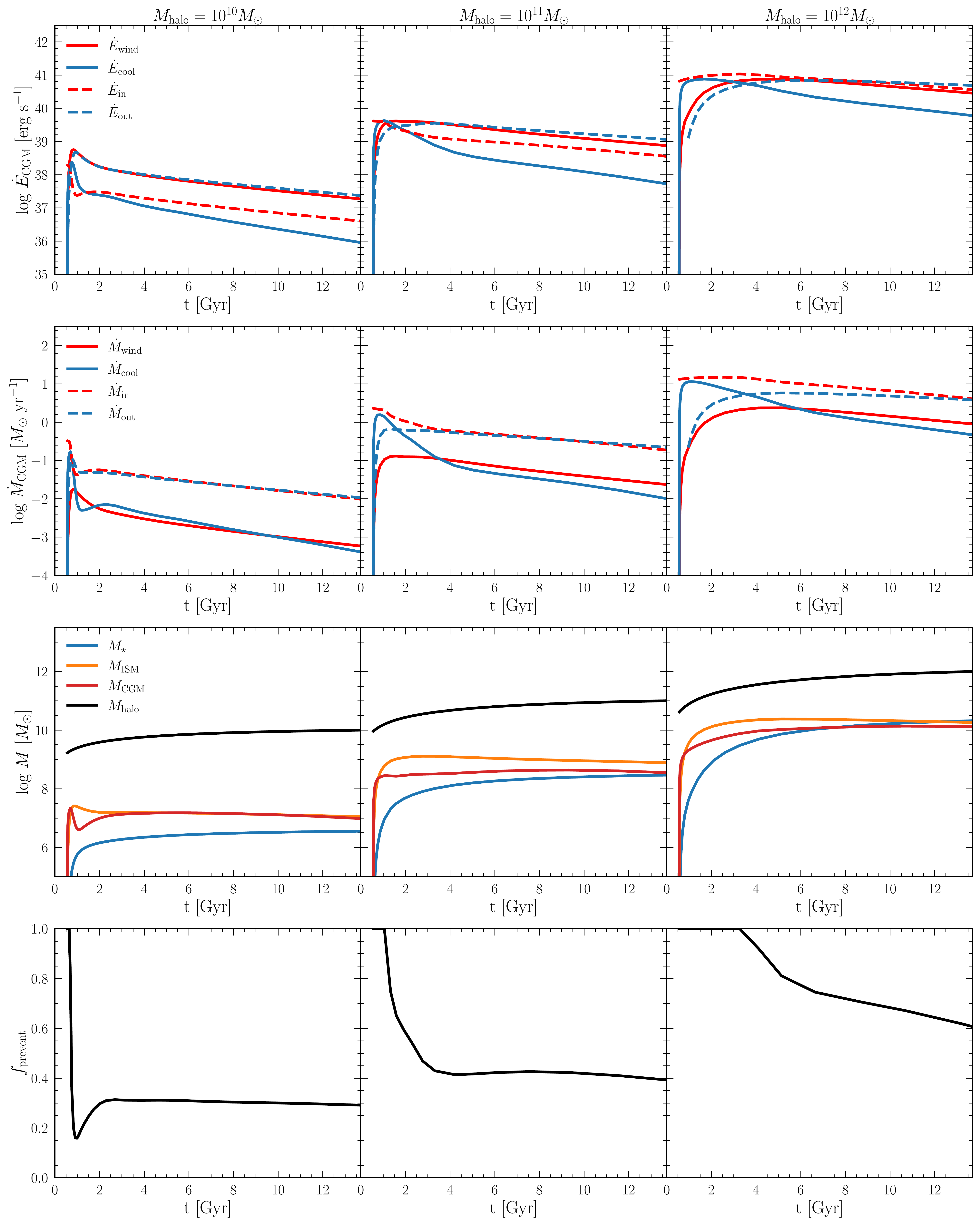}
    \caption{Time evolution of the logarithmic values of the flows in energy (\textbf{topmost row}) measured in ergs Gyr$^{-1}$ and mass flows (\textbf{middle-top row}) in $M_{\odot}$ Gyr$^{-1}$ for each sink and source term for the CGM reservoir in halos of mass $M_{\rm halo} = 10^{10} M_{\odot}$ (\textbf{left}), $10^{11} M_{\odot}$ (\textbf{center}), and $10^{12} M_{\odot}$ (\textbf{right}) respectively. Positive source terms are colored \textbf{red} while negative sink terms are in \textbf{blue}. \textbf{Solid} lines distinguish mass and energy flows at the galaxy-CGM interface, while \textbf{dashed} lines represent flows across the boundary of the CGM and IGM. The \textbf{middle-bottom row} tracks the time evolution of the mass of each reservoir in the regulator model, and the \textbf{bottom row} tracks the evolution of the preventative inflow parameter, $f_{\rm prevent}$, for each halo mass to the present-day.} 
    \label{fig: models}
\end{figure*}

\subsection{How the Model Self-Regulates}
The insensitivity we see to varying the mass loading factor of galactic outflows suggest that the model is self-regulated to a significant extent, by which we mean that changes to the composition of outflows prompts a change in the cooling efficiency of the CGM, which in turn shapes the properties of the galaxy through regulating future gas accretion. Increasing $\eta_M$ serves to enhance future accretion by increasing the density of circumgalactic gas, shortening the cooling timescale and enabling the rapid precipitation of gas back into the ISM in short order. The timescale of this re-accretion of gas in real galaxies is not well-understood, but likely could persist for multiple cycles, providing a significant source of gas accretion to galaxies at low redshift \citep{2010MNRAS.406.2325O, Ford2014, Angles-Alcazar2017}. Our model assumes the CGM to be well-mixed, supporting the picture of cool gas precipitating out of a hot halo \citep{Fraternali_hail}, where mass-loaded material from galactic winds can trigger condensation in excess of the amount of gas delivered to the CGM in outflows. 

This form of self-regulation is not the case for $\eta_E$; instead increasing $\eta_E$ enhances the heating of the CGM for a given amount of star formation, allowing more mass to be lifted out of the CGM, which prevents subsequent accretion by increasing $t_{\rm cool}$ of the CGM. What makes the model so sensitive to the energy-loading of the winds is its ability to remove baryons from the halo entirely, while adjusting $\eta_M$ or $\eta_Z$ only serves to shift baryons to a different component of the galaxy for a short while before they return back to the galaxy. We note that, in real galaxies, material unbound from the halo by energy-driven winds may eventually cool and return back to the halo on some (potentially long) timescale. Building a better understanding of both the timescales of recycled gas from outflows and the return of gas that has been expelled beyond the viral radius are required to better link observed galaxy scaling relations and the properties of SNe-driven winds. 

To understand how this self-regulation works in detail, we examine Fig.~\ref{fig: models}, which shows the energy and mass flows in three halos spanning the mass range of interest (with $10^{10} M_{\odot}$, $10^{11} M_{\odot}$, and $10^{12} M_{\odot}$ in the left, center and right columns, respectively). In each case, there is an initial period of adjustment between rapid cooling and heating by early star formation while the system comes into a quasi-equilibrium (on a timescale dictated by the depletion time) and then a generally slow evolution of both the energy and mass flows over time, with the evolution timescale set mostly by changes in the accretion rate (i.e. on a Hubble time). For the two lower mass halos, equilibrium is largely dictated by a balance between the energetic input from supernovae and the energy required to lift mass out of the CGM. Indeed, from the close balance of the cosmic inflow and CGM outflow terms in the $\dot{M}_{\rm CGM}$ curves (second row of panels), we see that the star formation rate is being set by the requirement that the CGM mass be held roughly constant (so that its cooling rate provides just enough mass to power the star formation). This simple balance of two terms is more complicated in the highest mass ($10^{12} M_{\odot}$) halo as the energy provided by the inflowing gas starts to become important (this is driven by the fact that the specific energy of the inflowing gas scales as $M^{2/3}$); however, the self-regulated quasi-equilibrium behaviour still holds.  

The result of this self-regulation can be seen in the bottom two rows. The mass builds up relatively slowly and energetic preventative feedback is working on two levels: regulating the CGM mass but also preventing the inflow of gas from the IGM. We note that the details of this evolution depend somewhat on the redshift evolution of the depletion time, which is not well constrained at high redshift (particularly for low mass halos). Our default selection is a relatively rapid evolution of $t_{\rm dep}$ such that the star formation is high at high redshift and the ISM mass is kept low; a more gradual evolution results in a slower build-up of the stellar component and a high ISM mass at high redshift. However, the overall self-regulatory behaviour is robust to changes in the depletion time, with the exception that longer depletion times at early times cause a longer period of oscillatory self-adjustment and, in extreme cases, can cause a large fraction of the overall gas accretion into the ISM to occur at early times, particularly for low-mass halos.

\begin{figure*}
	\centering
  	\includegraphics[width=1\textwidth]{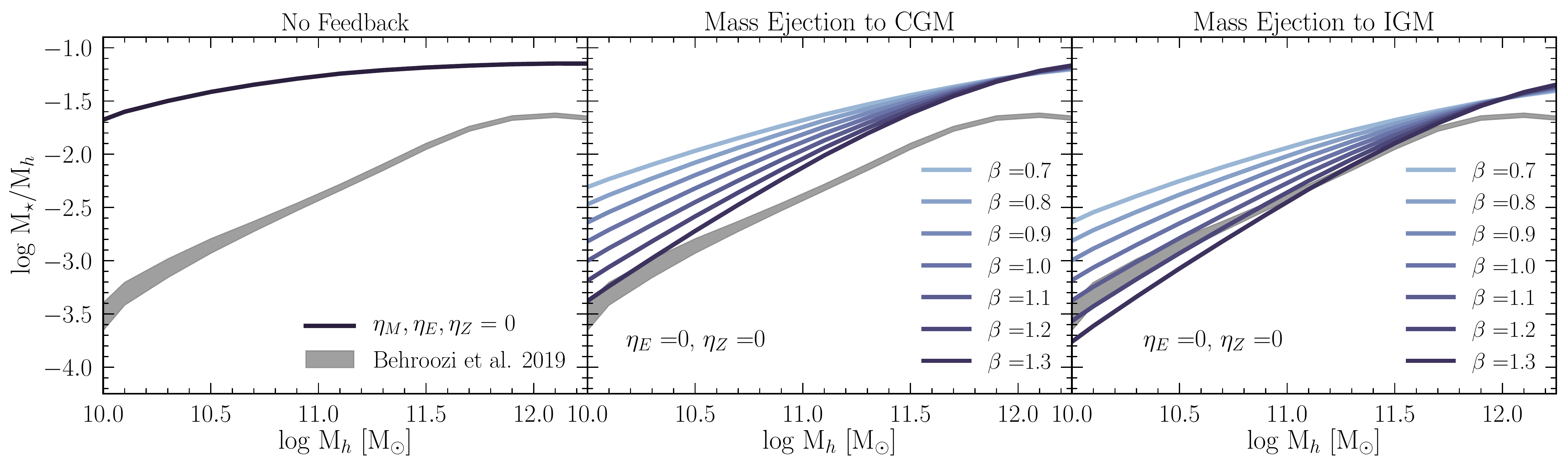}
    \caption{Log of $M_\star / M_{\rm h}$ from the regulator model as a function of halo mass, compared to the median observed stellar mass with uncertainties in gray and their derived halo mass from abundance matching from \citep{2019_behroozi}. The \textbf{left} plot shows the estimates of log of $M_\star / M_{\rm h}$ in the "No Feedback" case, where $\eta_M = \eta_E = \eta_Z = 0$. The \textbf{middle} and \textbf{right} plots show the same relation, but in the altered scenarios where winds of increasing mass-loading eject all of their mass to the CGM or mass is ejected directly to the IGM respectively, while $\eta_E =0$. The annotation of each plot displays which parameters were kept constant in the iterations of the model on display.}
    \label{fig: etae0}
\end{figure*}

\subsection{Importance of Preventative Feedback Compared to Ejective Feedback}
The leftmost panel of Figure \ref{fig: etae0} displays what happens to the $M_\star / M_{\rm halo}$-$M_{\rm halo}$ relation when we remove all feedback$-$setting $\eta_M$, $\eta_E$, and $\eta_Z$ to equal 0. We recover the classical result from early galaxy formation models of overcooling the baryons into the center of  the dark matter halo and the over-abundant production of stars \citep{white_core_1978, dekel_origin_1986, white_galaxy_1991}. 

Removing gas from the galaxy through galactic outflows with high mass-loaded winds has been a key part of most previous solutions to this problem \citep[e.g.,][]{keres_galaxies_2009}. To demonstrate that we can recover this result, that ejective feedback ($\eta_M > 0$) alone is capable of addressing the problems of overcooling, we set $\eta_E = 0$, removing the primary source of preventative feedback within the regulator model. We explore two instances where $\eta_E = 0$: one where all mass ejected from the central galaxy is deposited into the CGM and remains within the circulation of baryon flows (middle), and the extreme (but traditional) scenario where SNe-driven outflows launch all their gas out of the ISM such that it escapes to the IGM (right)

When ejected gas is deposited into the CGM with our simple power-law dependence of $\eta_M$ with halo mass, mass-loadings in low-mass galaxies only begin to touch the \cite{2019_behroozi} constraints when $\beta$ approaches $1.3$. This occurs because more efficient mass-loading is able to slightly reduce the stellar mass, but in the absence of a means to remove baryons from the halo, most gas recently ejected from the galaxy accretes back to the ISM in short-order and provides fuel for star formation. In the scenario where gas is driven directly out of the galaxy and does not return, we can significantly reduce the stellar mass, but large values for $\beta$ in eq.\ref{eq: eta_M} are required to drive down the stellar mass such that it agrees with observed galaxies, requiring mass-loadings as high as $\eta_M \sim 100$ in the lowest mass halo explored by our model, similar to values in previous cosmological simulations and SAM models \citep[e.g.,][]{2015_muratov, Angles-Alcazar2017, 2018MNRAS.473.4077P, Nelson2019,Pandya2020_SMAUG, Pandya_fire2}.

The difference between the two scenarios in the middle and left panels of Figure \ref{fig: etae0}, SNe-driven winds ejecting gas into the CGM or ejecting all of the gas into the IGM, demonstrate the prominent role of the CGM in regulating the baryonic content of galaxies through cooling and accretion. High specific energy feedback provides a means of preventing further accretion by heating the gas and providing additional thermal support against radiative losses, suppressing cosmic infall from the IGM, and lifting heated gas out of the halo.

\subsection{Connection to precipitation: the $t_{\rm cool}/t_{\rm ff}$-$M_{\rm halo}$ relation} \label{tctff} 
There has been considerable evidence from observations and simulations of the intracluster medium (ICM) that precipitation-regulated feedback from a central heating mechanism of the cluster suspends the ICM gas in a state of critical balance with the ratio of the cooling time to the freefall time $\gtrsim 10$ in massive galaxies \citep[for a recent review, see][]{Donahue2022}. Gas in the ICM becomes increasingly susceptible to condensation as the ratio falls below this precipitation limit. When this occurs, the condensing gas will fuel accretion onto the cluster central black hole and the resulting energetic feedback from the black hole (in the form of a jet) will heat the gas and lengthen the cooling timescale of the gas, bringing the circumgalactic gas back to equilibrium such that the ratio exceeds 10. In clusters, this process is thought to be the primary mechanism which keeps the galaxy quenched and the ICM regulated.


\begin{figure}[h!]
	\centering
  	\includegraphics[width=1\linewidth]{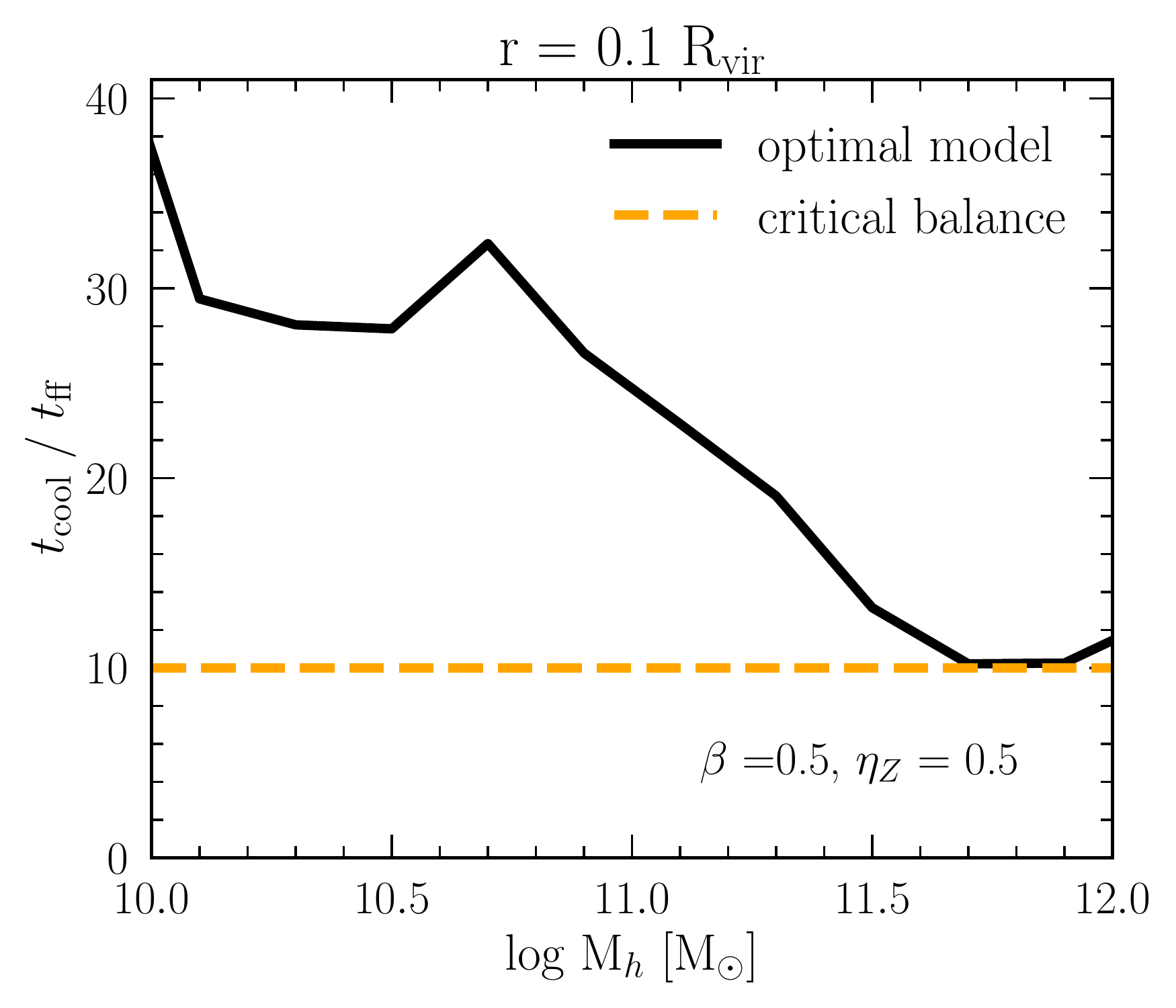}
    \caption{The ratio of the cooling time to the freefall time, $t_{\rm cool}/t_{\rm ff}$, evaluated at $R = 0.1 R_{\rm vir}$ as a function of halo mass with the optimal choice of $\eta_E$ from recursive bisection. The annotations on the bottom plot display which parameters were kept constant. The orange dashed line marks critical balance, the predicted $t_{\rm cool}/t_{\rm ff}\gtrsim 10$ of gas in precipitation models of the CGM.}
    \label{fig: tctff}
\end{figure}

This same mechanism has been suggested to be operating in galaxies, with the ICM replaced by the CGM, and SN playing the role of the central black hole \citep{Voit2015b}. \cite{Fielding2017} showed this mechanism could be at play in the CGM when $\eta_E/\eta_M$ is sufficiently high. The model we described in this paper is similar in many respects but is not directly connected to precipitation (although that is a natural manifestation of our CGM cooling). In this section, we explore the connection to see if the CGM properties we find can be matched on to this critical balance picture.

The cooling time extends to longer timescales as the density drops farther out in the halo, while the freefall time of the gas also increases with distance (generally more slowly), giving the ratio a slowly rising radial profile. For Figure \ref{fig: tctff}, we choose to evaluate the $t_{\rm cool}/t_{\rm ff}$ ratio as a function of halo mass at a common radius of 0.1$R_{\rm vir}$ for our optimal model for the energy-loading found in section \ref{sec: bisect}. Generally, if we evaluated it at small radii, the ratio would be lower (although this depends on the density and temperature profiles we adopt).

We find that $t_{\rm cool}/t_{\rm ff}$ is somewhat larger than the cannonical precipitaiton prediction in low mass galaxies, approaching values of $\sim 30-40$, but this ratio drops with increasing halo mass, reaching ratios close to $t_{\rm cool}/t_{\rm ff} \simeq 10$ for Milky Way mass halos. The long cooling times in the low-mass halos seen in our model arise because of the high energy-loading (and low mass-loading) in the galactic winds from supernovae, which raises the CGM temperature well above the virial temperature and makes the CGM strongly over-pressurized, as well as lowering the CGM density due to high mass and energy outflows out of the halo. The amount of energy and mass lifted out of the CGM is sensitive to what we assume for the specific energy of these outflows, which is not well understood. For simplicity we assume that the specific energy of these halo outflows is equal to the average specific energy of the entire CGM, but this need not be the case in real galaxies, where it is possible that the gas lifted out of the halo could have a much higher specific energy. Such outflows would lower the average temperature in the CGM and potentially bring the remaining gas closer to the precipitation picture than what our model predicts for low mass halos. 

Our ratios for $t_{\rm cool}/t_{\rm ff}$ can also be compared with certain cooling flow models of the CGM, where ineffective heating allows cool gas to cascade inward on a cooling timescale \citep{Stern2019}. Such models for a Milky Way halo also find ratios between $t_{\rm cool}$ and $t_{\rm ff}$ that approximately equal $\sim 7-10$ when gas accretion and the star formation are approximately equal. Such cooling flow models also predict shorter cooling times in lower mass halos, which is in contrast to our model, where energy outflows from supernovae in these galaxies is very effective at heating the CGM and constraining gas accretion. 

\subsection{Comparison to Other Regulator Models}
Similar analytic regulator models in the past have also considered treatments of preventative and ejective feedback. Although they do not explicitly solve for the CGM reservoir, \cite{dave_analytic_2011} included a preventative feedback parameter to quantify the amount of gas that enters the halo but is prevented from reaching the ISM. This preventative feedback includes not only contributions from winds, but also photoionization, quenching, and virial shocks. Although their inclusion of preventative feedback is largely illustrative, they do converge onto a picture of preventative feedback being a stronger contributor that limits gas accretion in lower mass galaxies ($\sim 10^{10} M_{\odot}$) and declining in importance for more massive halos). 

Key differences between our regulator model and past constructions lay in determining how feedback processes set the stellar-to-halo mass relation and its low-mass end slope. \cite{lilly_gas_2013} find with their gas-regulator model that $\sim 40 \%$ of baryons that flow into the halos of galaxies across a stellar mass range from $10^{9}$ to $10^{11} M_{\odot}$ are processed within galaxies. Their fit to SDSS data from \cite{Mannucci2010} yield $\eta_M$ values of $0.2-0.3$ for $M_{\star} = 10^{10} M_{\odot}$ systems, which are comparable to values assumed in this work for similar systems, but this agreement would not be true for lower-mass systems due to the steep inverse dependence between mass-loading factor and stellar mass required in their fit to match the data. The equilibrium model used in \cite{Mitra2015}, employing a similar preventative feedback scheme to \cite{dave_analytic_2011} and fitting $\eta_M$ to galaxy scaling relations, find $\eta_M \propto M_{\star}^{-0.5}$, not much larger than our $\beta=0.5$ scaling, however progressively larger mass outflows are required to match observations at higher redshift. More recent work from \cite{Mitchell2022} build a regulator-like model based on reproducing the flow terms from the EAGLE cosmological simulation, finding ejective feedback out of galaxies and halos to be the primary process shaping the stellar-to-halo mass relation, although with preventative feedback also playing a significant role. High mass loadings, particularly in low-mass galaxies, are a persistent feature of most galactic regulator models, which makes our model, accounting for the energy content of the CGM and self-consistent treatment of preventative feedback, a significant departure from past regulator models.

Other preventative feedback models have placed the location of galaxy regulation in the IGM, where pre-heating in the IGM prevents a portion of baryons from collapsing into halos due to their enhanced thermal pressure \citep{2015_Lu_preventative}. These models rely predominantly on preventative feedback with low mass-loading at late times, and have also found some success in matching certain galaxy scaling relations, such as the size evolution of galactic discs \citep{2015_Lu_disc-formation} and the mass-metallicity relation of low-mass galaxies \citep{Lu2017}. The role each of these preventative mechanisms named above (or others) play in regulating star formation and their interactions with one another remains largely unexplored territory in galactic modeling.

In our model of the energy regulation of the CGM, we left out non-thermal contributions to the energy, but the CGM is also home to a significant kinetic energy reservoir. This kinetic energy may manifest as turbulence or bulk flows which may provide their own form of pressure support against gravity in the CGM that is not entirely thermal \citep{Fielding2017}. Incorporating this kinetic energy component will be the subject of a future paper (Pandya et al., in prep.). In addition, cosmic rays can potentially operate as a preventative mechanism in the CGM, but we leave this for later work.

\section{Summary \& Conclusion} \label{sec: 5}

Wind outflows driven by supernovae are a powerful mechanism for regulating the baryonic contents of galaxies. We adopt a simple gas regulator model to understand how different mass and energy wind loading factors affect the global properties of galaxies and their scaling relations. We include the CGM as a reservoir; adopting a simple radial density profile, and track the flows of mass, metallicity, and energy, along with the gas mass and stellar mass of the central galaxy. We run the regulator model over a halo mass range of $\log{M_{\rm halo} / M_{\odot}} = 10$ to 12 and compare the galaxy properties at the present-day to a set of galaxy scaling relations. The major conclusions of our work are described below:
\begin{itemize}[itemsep=1mm, parsep=0pt]
    \item A gas-regulator model tracking the flows of mass, metallicity, and energy in and out of the CGM reservoir is able to qualitatively reproduce the broad features of galaxy scaling relations such as the $M_{\star}/M_{\rm halo}-M_{\rm halo}$ relation and the $M_{\rm ISM}/M_\star - M_\star$ relation over the range $\log{M_{\rm halo} / M_{\odot}} = 10$ to 12, where SNe are believed to be the dominant source of baryonic feedback.
    \item The estimated stellar mass near the present-day redshift is robust against choices for $\eta_M$ and $\eta_Z$, but is sensitive to the value of $\eta_E$, suggesting that preventative feedback from energetic outflows could play a significant role in limiting subsequent gas accretion into the ISM. High energy-loaded winds heat the CGM and lift heated gas out of the halo, reducing the density of the CGM and suppressing cosmic infall from the IGM. This limits the cooling efficiency of the CGM and accretion to the ISM. The $M_{\star}/M_{\rm halo}-M_{\rm halo}$ relation favors increasing values of $\eta_E$ for decreasing $M_{\rm halo}$, where $\eta_E \sim 1$ for $10^{10} M_{\odot}$ mass halos and $\sim 0.1$ for Milky Way-like halos.
    \item The CGM plays a significant role in regulating mass outflows from the galaxy when the CGM is overpressurized. Highly mass-loaded outflows from the galaxy increase the mass and density of the CGM and lower the CGM's specific energy, inhibiting its ability to suppress cosmic inflow from the IGM. These processes enhance future accretion from the CGM through shortening the cooling timescale and enables the rapid accretion of gas back into the ISM. 
    \item Our model of a hot CGM with a radial density profile going as $\propto r^{-1.4}$ produces mass estimates of the CGM that are broadly consistent with observational constraints from the Milky Way. The mass and metallicity of the CGM is strongly linked to the composition of galactic outflows. 
    \item Larger values of $\eta_M$ increase the mass left over in the CGM reservoir by the present-day, whereas greater $\eta_E$ and $\eta_Z$ diminish the CGM mass through lifting heated gas out of the halo and enhancing accretion from radiative cooling respectively. 
    \item SNe-driven winds regulate the cooling efficiency of the CGM, and the relation between the cooling time, $t_{\rm cool}$, and the free-fall time, $t_{\rm ff}$ of circumgalactic gas. When evaluating $t_{\rm cool}/t_{\rm ff}$ at a common radius of 0.1 $R_{\rm vir}$, the value of the $t_{\rm cool}/t_{\rm ff}$ is largest in the CGM of low-mass halos, reaching as high as $30-40$ due to very effective heating from supernovae which raises the CGM temperature well above the virial temperature. The ratio $t_{\rm cool}/t_{\rm ff}$ decreases with increasing halo mass and approaches the predicted ratio from precipitation and cooling flow models of $\approx 10$ in Milky Way-like halos. 
    \item Feedback through strong mass ejection from the ISM alone is unable to reproduce the $M_{\star}/M_{\rm halo}-M_{\rm halo}$ relation in the scenario if material ejected from the galaxy is deposited to the CGM, unless mass-loadings in low-mass galaxies are considerable larger than those considered in this work. It is sufficient however in the (more extreme) scenario in which the ejected gas is delivered directly to the IGM.
\end{itemize}

\begin{acknowledgements}

GLB acknowledges support from the NSF (AST-2108470, XSEDE grant MCA06N030), NASA TCAN award 80NSSC21K1053, and the Simons Foundation (grant 822237). Support for this work was provided by NASA through the NASA Hubble Fellowship grant HST-HF2-51489 awarded by the Space Telescope Science Institute, which is operated by the Association of Universities for Research in Astronomy, Inc., for NASA, under contract NAS5-26555. RSS is supported by the Simons Foundation. 
\end{acknowledgements}

\bibliography{Winds}
\bibliographystyle{aasjournal}

\appendix
\restartappendixnumbering
\section{Depletion Time Sensitivity}  \label{app: A}
Here we test the sensitivity of our regulator model to assumed parameters outside our characterization of wind outflows. Firstly, the depletion time sets the timescale for the conversion of gas into stars. This conversion is observed to be more efficient in Milky Way-mass galaxies (i.e. higher stellar masses). This is expressed in eq.~\ref{eq: tdep} used for $t_{\rm dep}$ referenced in section \ref{sec: star_m}. We can modulate the normalization of $t_{\rm dep}$, and its power-law dependence. In prior sections, we use the depletion time estimate derived from \cite{McGaugh_mainsequence_2017}, where the logarithmic depletion time is normalized by a prefactor equal to 4.92 and and a dependence on stellar mass with a power law index of 0.37. We define the normalization of depletion time as a parameter called $C$ and its power law index as $\gamma$. For this section, we keep one parameter set to its fiducial value, while varying the other. Since the influence of $t_{\rm dep}$ should primarily cover the conversion of gas into stars in the galaxy, we place a special emphasis on the $M_\star / M_{\rm halo}$-$M_{\rm halo}$ relation, and the $M_{\rm ISM}/M_\star$-$M_\star$ relation. For these tests and those in Appendix \ref{app: B}, we use the power law formula dependent on halo mass introduced in eq \ref{eq: eta_M} with a power law index of $\beta = 0.5$, the power law fit of $\eta_E$ from eq. \ref{eq: eta_E} and its best-fit values, and set $\eta_Z = 0.2$.

\begin{figure*}[h!]
	\centering
  	\includegraphics[width=0.7\textwidth]{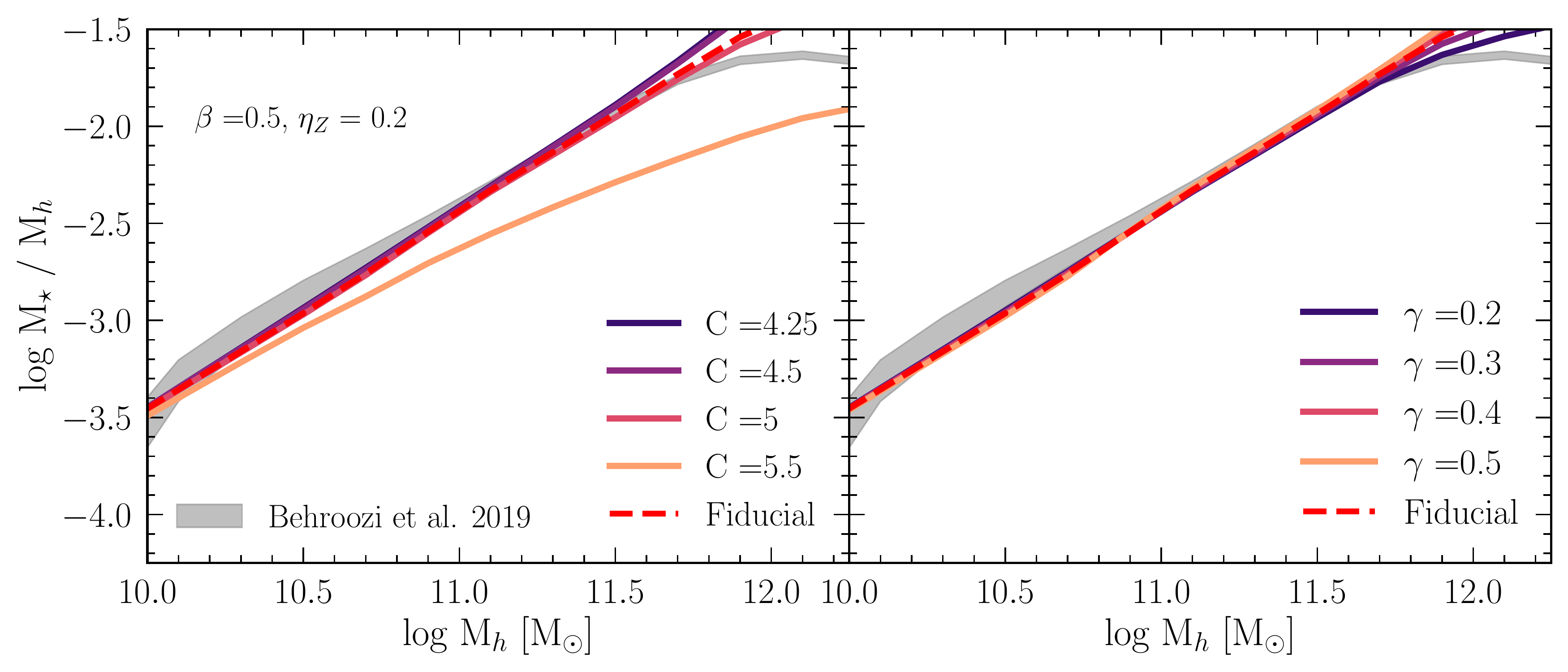}
    \caption{Log of $M_\star / M_{\rm h}$ from the regulator model as a function of halo mass, compared to the median observed stellar mass with uncertainties in gray and their derived halo mass from abundance matching from \citet{2019_behroozi}. Estimates of $M_\star / M_{\rm halo}$ are presented as a function of the normalization of the depletion time $C$ (\textbf{left}), and its power-law dependence, $\gamma$ (\textbf{right}).}
    \label{fig: MsMh_tdep}
\end{figure*}

\begin{figure*}[h!]
	\centering
  	\includegraphics[width=0.7\textwidth]{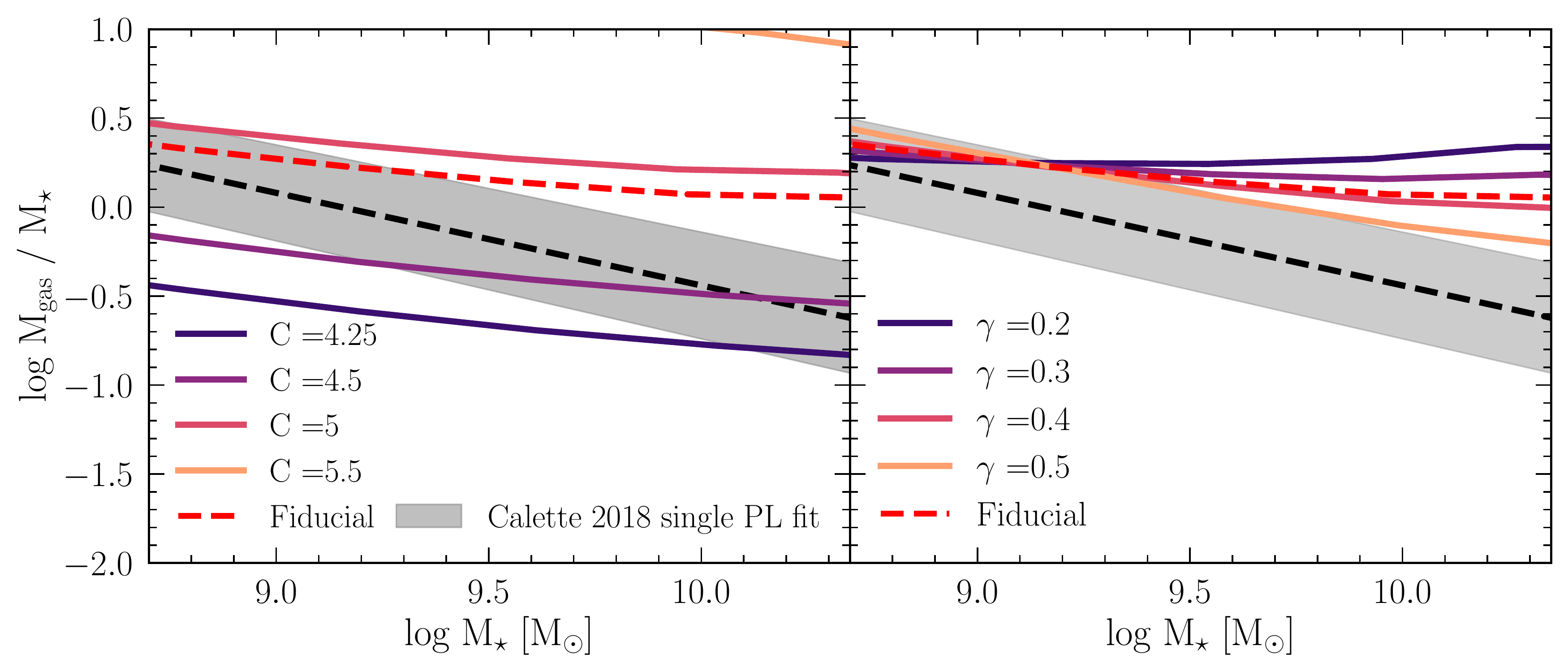}
    \caption{Log of $M_{\rm ISM} / M_\star$ from the regulator model as a function of stellar mass, compared to the best-fit single power-law fit for the total cold gas mass as a function of $M_\star$ over the range $7.3\lesssim\log{(M_\star / M_{\rm h})}\lesssim11.2$ for late-type galaxies from \citet{calette_hi-_2018}. Shaded lines correspond to alterations in the normalization of the depletion time $C$ (\textbf{left}) and its power-law dependence with stellar mass, $\gamma$ (\textbf{right}) and their respective influence on the $M_{\rm ISM}-M_\star$ relation.}
    \label{fig: MgMs_tdep}
\end{figure*}


We show in the left-hand plot of Figure \ref{fig: MsMh_tdep} that modulating the normalization of $t_{\rm dep}$ over an order of magnitude from $C=4.25-5.5$ does have a noticeable effect on the $M_{\star}/M_{\rm halo}$-$M_{\rm halo}$ relation, such that larger (smaller) values result in a reduced (greater) stellar masses, especially in more massive halos. This is partly to be expected, considering that a larger normalization would lead to larger depletion timescales and a reduced star formation rate. The less efficient conversion of gas into stars leads a greater proportion of gas to stars in all galaxies as a function of halo mass, manifesting as a vertical shift in the $M_{\rm ISM}/M_\star$-$M_\star$ relation of the model (left-hand plot of Figure \ref{fig: MgMs_tdep}).


The right-hand plot of Figure \ref{fig: MsMh_tdep} shows how modifying the power-law index of $t_{\rm dep}$ has very little effect on the $M_{\star}/M_{\rm halo}$-$M_{\rm halo}$ relation. The most apparent consequence of modifying $\gamma$ in seen in Figure \ref{fig: MgMs_tdep}, where it is shown that the power law index of the depletion time alters the slope of the $M_{\rm ISM}/M_\star$-$M_\star$ relation produced by the model. Increasing the index reduces the star formation efficiency in low-mass halos and increases its efficiency in more massive halos and vice-versa for its reduction.  Larger values of $\gamma$ are in better agreement with the slope of the PL fit from \cite{calette_hi-_2018}, with $\gamma \geq 0.5$ producing the closest match with their observational constraints. 


The robustness of the stellar mass to the power law dependence of the depletion time and even its normalization at the lower mass end in Figure \ref{fig: MsMh_tdep} may point to the important role that $t_{\rm dep}$ plays in setting the timescale for self-regulation to arise at high redshift. Longer depletion times, especially in low mass halos, delay the ignition of star formation and the high specific energy outflows which prevent accretion into the CGM and the ISM. This means that although star formation in the disk becomes less efficient, the increased baryon fraction in the halo leaves the resulting stellar mass largely unchanged. Shorter depletion times in low-mass halos (lower power-law index) means that they are more efficient in their star production, but suppress cosmic inflow through self-regulation much earlier, leaving less gas in the galaxy to be processed into stars.

\section{Density Profile Sensitivity} \label{app: B}

\begin{figure}
	\centering
  	\includegraphics[width=0.5\linewidth]{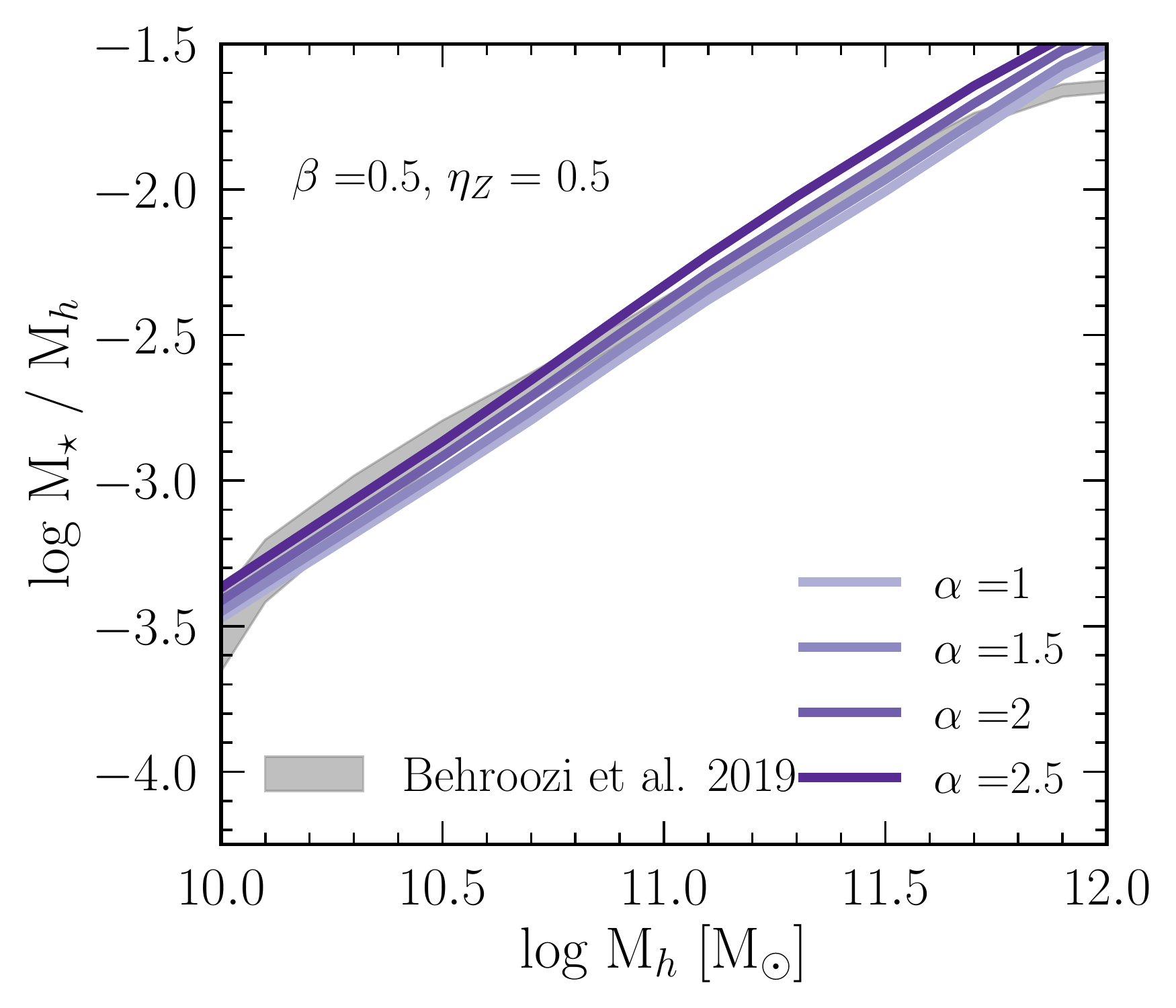}
    \caption{Log of $M_\star / M_{\rm h}$ from the regulator model as a function of halo mass, compared to the median observed stellar mass with uncertainties in gray and their derived halo mass from abundance matching from \citep{2019_behroozi}. Shaded lines represent different variations of the power-law index for the CGM radial density profile, $\alpha$, and its effect on the $M_\star / M_{\rm halo}$-$M_{\rm halo}$ relation.}
    \label{fig: MsMh_a}
\end{figure}

Another important quantity in our model is our assumed density profile of the CGM. In eq.(\ref{eq:density}), we assume a power-law dependence with distance from the center with an index of $\alpha=1.4$. Since the density profile of the CGM is not well-constrained, it is important to observe how our models responds to different assumptions of the gas density profile. Varying $\alpha$ in our parameterization of the density profile changes how $\rho(r)$ depends on radius, where larger values of $\alpha$ produce a steeper density profile where a greater concentration of gas is placed in the inner CGM relative to its outskirts. 

\begin{figure}
	\centering
  	\includegraphics[width=0.5\linewidth]{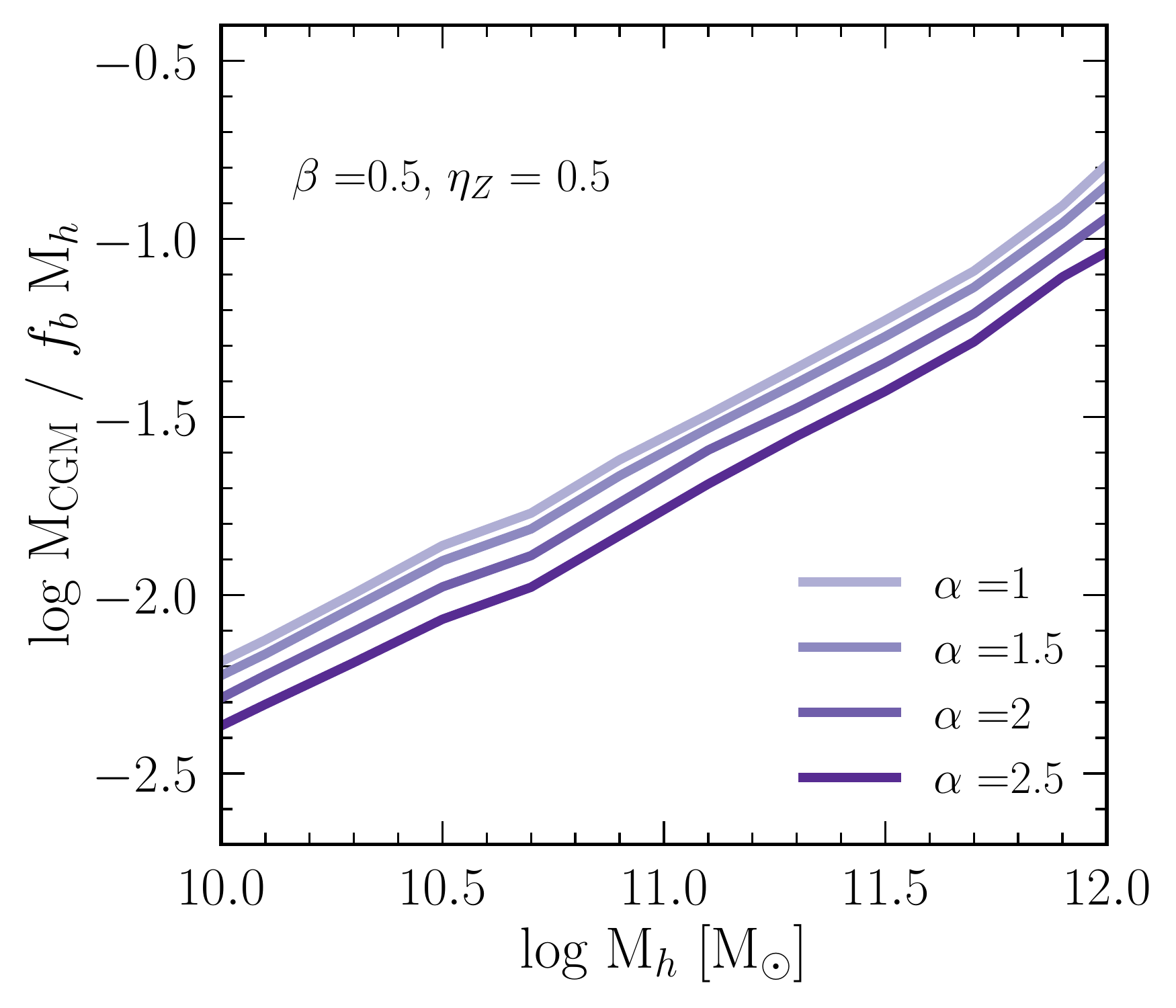}
    \caption{Estimates of $\log(M_{\rm CGM} / M_{\rm h})$ as a function of halo mass. Estimates of $M_{\rm CGM} / M_{\rm halo}$ are presented as a function of the power-law index for the CGM radial density profile, $\alpha$.}
    \label{fig: Mcgm_a}
\end{figure}

After plotting the $M_\star / M_{\rm halo}$-$M_{\rm halo}$ relation for varying forms of the CGM density profile in Figure \ref{fig: MsMh_a}, we find that larger values of $\alpha$ increase the stellar mass at $z=0$ for all halos. With a greater concentration of mass in the inner CGM, steeper density profiles elevate the overall cooling efficiency of gas near the galaxy and result in greater accretion and more fuel for star formation. This rise in star formation should also work to increase the total outflow of energy into the CGM, but this countering response to enhanced accretion is not enough to fully oppose the significant radiative losses in the inner gas halo. Steeper density profiles have a predictable effect on the CGM. In Figure~\ref{fig: Mcgm_a}, larger values of $\alpha$ result in less mass in the CGM at late times due to the greater accretion from the CGM to the ISM.



\end{document}